\documentclass{article}

\usepackage{amsmath}
\usepackage{graphicx}
\usepackage{cite}
 
\input epsf

\begin{document}

\setlength{\unitlength}{1mm}

\begin{titlepage}

\begin{flushright}
Edinburgh 2001/19\\
LAPTH-873/01\\
November 2001
\end{flushright}
\vspace{1.cm}

\begin{center}
\large\bf
{\LARGE \bf A NEXT-TO-LEADING ORDER STUDY OF PION-PAIR 
PRODUCTION \& COMPARISON WITH E706 DATA}\\[2cm]
\rm
{ T.~Binoth$^{b}$, J.~Ph.~Guillet$^{a}$, E.~Pilon$^{a}$ and M. Werlen$^{a}$}
\\[.5cm]

{\em $^{a}$ Laboratoire d'Annecy-Le-Vieux de Physique Th\'eorique LAPTH,}\\
{\em Chemin de Bellevue, B.P. 110, F-74941  Annecy-le-Vieux, France}\\[.2cm]
{\em $^{b}$ Department of Physics and Astronomy,
University of Edinburgh, }\\
{\em  Edinburgh EH9 3JZ, Scotland}        
      
\end{center}
\normalsize
\vspace{2cm}

\begin{abstract} 
\noindent
We discuss the production of pion pairs with a large invariant mass in hadronic
collisions. We present a study based on a perturbative QCD calculation at full
next-to-leading order accuracy, implemented in the computer programme {\em
DIPHOX}. We provide a comparison of various predictions with the  corresponding
observables measured by the E706 fixed target experiment. Discrepancies between data and next-to-leading order calculation are carefully analysed. We classify the theoretical
next-to-leading order distributions with respect to their infra-red sensitivity,
and explain which distributions need improvements. Further, we comment on the
energy scale dependences of non pertubative effects.
\end{abstract}



\end{titlepage}

\section{Introduction}\label{intro}

Quantum Chromodynamics (QCD) is now accepted as the theory of the strong
interaction. With the starting run II of the Fermilab Tevatron, and the
forthcoming Large Hadron Collider (LHC) at CERN, QCD is entering a new area
associating improved precision measurements with more accurate calculations.
This concerns not only a refined study of the strong interaction itself and, in
particular, the ability of perturbative QCD to describe high energy phenomena
precisely, but also the accurate prediction of sensitivities in Higgs
boson and  expected new physics searches. Both need an accurate understanding
of respective signals and backgrounds, which are heavily affected by QCD
radiative corrections. Meanwhile, the proper understanding of the large amount
of data collected by fixed target experiments remains rather challenging. In
all these respects, the production of neutral pion pairs with a large invariant
mass is an interesting phenomenon. This process has been recently measured by
the fixed target E706 Experiment \cite{e706} at the Fermilab Tevatron. The
study of this process and the examinations of the various observables which
have been measured are instructive for several reasons.

\vspace{0.2cm}

\noindent
First of all, it is interesting to compare the ability of perturbative QCD to
describe single pion and pion pair production respectively. In particular, one
can identify the observables for which non perturbative effects are negligible
and the QCD improved parton model based on collinear factorisation can be
safely used, and examine otherwise when it fails and why. In this respect,
loosely speaking, inclusive enough correlation observables with symmetric
selection cuts are less sensitive to non perturbative effects such as, e.g. a
$k_{T}$ kick, than one particle inclusive distributions
\cite{baier-engels-peterson}, especially at fixed target energies. Dipion
observables thus provide a good opportunity to cross-check the consistency of
the sets of parton-to-pion fragmentation functions $D_{i \to
\pi^{0}}(z,M^{2}_{f})$, such as those given in Ref.~\cite{bkk} and the more
recent ones in Ref.~\cite{kkp}, which are available in the
literature\footnote{Similar consistency checks for the set determined in
\cite{kkp}, especially for the gluon fragmentation functions, have been
presented in Ref.~\cite{kkp-univ} in the case of hadro- and photo-production of
single inclusive charged hadrons.}\label{f1}. Both these sets have been
extracted from data on hadronic final states in $e^{+}e^{-}$ annihilation. They
are given by analytical ans\"{a}tze whose parameters were fitted in the range
$0.1 \leq z \leq 0.9$ but which are tightly constrained\footnote{To be precise,
regarding pions, the authors of \cite{kkp} determine fragmentation functions
into charged pions then deduce the ones into neutral pions relying on isospin
symmetry.}\label{ftnt1} only in $0.1 \leq z \leq 0.7$. In particular, the
gluon-to-pion fragmentation functions are not much constrained because the
subprocesses involving gluons in $e^{+}e^{-}$ annihilation appear at higher
order in the perturbative expansion. Instead, subprocesses involving outgoing
gluons contribute already at lowest order in hadronic collisions. Therefore the
contributions from gluon fragmentation in hadronic collisions are not
negligible and large uncertainties on their magnitude may affect quantitative 
predictions. 

\vspace{0.2cm}

\noindent 
In addition, the production of $\pi^{0}$ pairs, together with the production
of  pairs $\gamma \pi^{0}$, where $\gamma$ is itself a prompt photon, is the
dominant background against which the experiments measuring prompt photon pairs
have to fight. The expression ``prompt photon" means that such a photon does
not come from the decays of a hadron such as $\pi^{0}$, $\eta$, etc. produced
at large transverse momentum. This  background is especially worrisome in
collider experiments, such as the Collider Detector at Fermilab (CDF)
\cite{cdf} and D0 \cite{d0} experiments at  the Fermilab Tevatron, where the
pions cannot be all reconstructed event by event. In addition, these two
processes of $\gamma \pi^{0}$ and $\pi^{0} \pi^{0}$ production represent an
important fraction of the so-called reducible background to the search of
neutral Higgs bosons in the channel $H \rightarrow \gamma \gamma$ in the
intermediate mass range 80 -- 140 GeV at the CERN Large Hadron Collider (LHC)
\cite{atlas,cms,atlas+cms,cernreplhc}. Since the jet-jet cross section is about eight
orders of magnitude larger than the expected Higgs boson signal for a standard
model Higgs boson, this reducible background is overwhelming before any
selection cut is imposed. Stringent isolation cuts are applied in order to
reduce this background severely and make prompt photons measurable.
Nevertheless, it is a theoretical challenge to understand the remaining
background at a quantitative level.  In this respect, the understanding of
$\pi^0$  pair production in the fixed target energy range will allow one to
transport this knowledge to collider energies \cite{bgpw2} using the
factorization property of perturbative QCD.

\vspace{0.2cm}

\noindent
In the present article, we present a phenomenological study of $\pi^{0}$ pair
production at Next-to-Leading Order (NLO) accuracy in perturbative QCD, based
on the computer code {\em DIPHOX} \cite{bgpw}. In section \ref{theory} we
briefly describe the theoretical content on which the present study is based.
We then compare {\em DIPHOX}-based theoretical results with data of the E706
experiment in section \ref{comparison-e706}. Section \ref{phenom} is then
dedicated to a detailed phenomenological discussion of the various observables
measured in $\pi^{0}$ pair production. The aim of this phenomenological study
is 1) to test the ability of the theoretical framework used to describe various
correlation observables that have been measured by the E706 experiment, 2) to
identify and understand the possible discrepancies between theory and data for
these observables, and 3) to examine how the sensitivity to non perturbative
effects, which limit the predictive power of the perturbative approach, evolves
as the `hardness' of the probes increases. To this purpose we classify the
various observables according to their sensitivity to the infrared dynamics. At
this occasion, the analysis of the transverse distribution of each pion leads
us to reconsider briefly the $p_{T}$ spectrum in the case of inclusive one pion
production as well. Section \ref{concl} gathers our conclusions and
perspectives. The goal is to  foresee how reliable predictions based on our
present knowledge can be for dipion production at much higher, collider
energies. In a forthcoming paper, we will investigate more specifically our
ability to predict the reducible background to the Higgs boson search at LHC in
the channel $H \to \gamma \gamma$, relying on the informations obtained in the
present study.

\section{Theoretical content}\label{theory}

The computer program {\it DIPHOX} on which the present study relies is a Monte
Carlo code of partonic event generator type. This code has been designed to
describe the production of pairs of particles in hadronic collisions,
accounting for all contributing partonic processes at full NLO accuracy. These
particles can be prompt photons or hadrons, in particular neutral pions. This
code is flexible enough to accommodate various kinematic or calorimetric cuts.
Especially, it allows to compute cross sections for isolated $\gamma \pi^{0}$
and $\pi^{0} \pi^{0}$  pairs, for any infrared  and collinear safe isolation
criterion which can be implemented at the partonic level. The physical content
and schematic description of {\it DIPHOX} has been given in \cite{bgpw}, where we have concentrated on the case of diphoton production\footnote{The acronym {\it
DIPHOX} stands for "diphoton cross sections".} \cite{new14,new15}. However, the production 
mechanisms of fragmentation type involved in diphoton production are exactly
the same as the ones at work in the hadroproduction of $\gamma \, \pi^{0}$ and
$\pi^{0} \pi^{0}$. Let us remind this point briefly.

\vspace{0.2cm}

\noindent
A prompt photon may be produced according to two possible mechanisms: either it
takes part directly to the hard subprocess (let us call it call ``direct"), or
it results from the collinear fragmentation of a parton which is itself
produced with a large transverse momentum (let us call it ``from
fragmentation"). Consequently, pairs of prompt photons with a large invariant
mass may be produced  according to three possible mechanisms: both of them take
part directly to the hard subprocess (direct mechanism), or one of them takes
part directly to the hard subprocess while the other one results from the
fragmentation of a large transverse momentum parton (one-fragmentation
mechanism), or else each of them results from fragmentation (two-fragmentation
mechanism). For a general discussion of the production mechanisms of prompt
photon pairs, see \cite{bgpw}. Similarly, $\gamma \, \pi^{0}$ pairs with a
large invariant mass may be produced according to two possible mechanisms:
either the prompt photon takes part directly to the hard subprocess (direct
mechanism), or it results from the collinear fragmentation of a parton which is
itself produced with a large transverse momentum (fragmentation mechanism). 

\vspace{0.2cm}

\noindent
In agreement with the factorisation property - even in the case of reasonable
isolation requirements, see Ref.~\cite{cfgp} - the partonic subprocesses
involved by the one-fragmentation mechanism in prompt photon pair
production are the same as the ones involved in $\gamma \, \pi^{0}$ production
when the prompt photon is of the `direct type'. The only change is the
fragmentation functions involved: fragmentation into a photon in the first
case, into a pion in the last one. Similarly, the partonic subprocesses
involved by the so-called two-fragmentation mechanism in prompt photon pair
production are the same as both the ones involved in $\gamma \pi^{0}$ 
production when the prompt photon come from fragmentation mechanism, and those
involved in $\pi^{0}$ pair production. Therefore, by keeping only the relevant
mechanisms, and replacing the fragmentation functions appropriately, {\it
DIPHOX} allows also to study the production of $\gamma \, \pi^{0}$ as well as
of $\pi^{0}$ pairs  with a large invariant mass and individual transverse
momenta, at full  NLO accuracy.

\section{Comparison with the E706 data}\label{comparison-e706}

The E706 experiment \cite{e706} studied collisions\footnote{The inclusive
hadroproduction of pion pairs has also been studied earlier by the CCOR ISR
experiment \cite{ccor} and the NA24 fixed target experiment \cite{na24} at
CERN. We choose to focus on the data collected by the
E706 experiment, which is more recent and gathers higher statistics.} on
various fixed targets (H,Be,Cu) of several beams. They used proton beams at
different beam energies, 530 GeV ($\sqrt{S} \simeq 31.55$ GeV) and 800 GeV
($\sqrt{S} \simeq 38.76$ GeV), as well as a 515 GeV $\pi^{-}$ beam ($\sqrt{S}
\simeq 31.10$ GeV). We have decided to focus on comparisons with data from
proton initiated reactions, because the sets of available pion's pdf are rather
old and not as reliable as the recent ones for protons. We have also focused on
data on Hydrogen and Beryllium targets, because the Beryllium nucleus is still
rather small hence nuclear effects are hopefully reduced, in contrast to the
case of a Copper target where nuclear effects are more important. We have
performed a comparison between our NLO QCD results and data on Hydrogen and
Beryllium targets, for protons beam energies of 530 GeV and 800 GeV. We have
chosen to present only the comparison with the data obtained in proton-Beryllium
for which the statistics is larger than for Hydrogen target. We restrict the
presentation mainly to a comparison with data from the collision of the 530 GeV
proton beam, in order to limit the proliferation of plots in the article.
Similar conclusions have been obtained in the other cases studied.

\vspace{0.2cm}

\begin{figure}[htbp]
\begin{center}
\includegraphics[width=0.8\linewidth,height=12 cm]{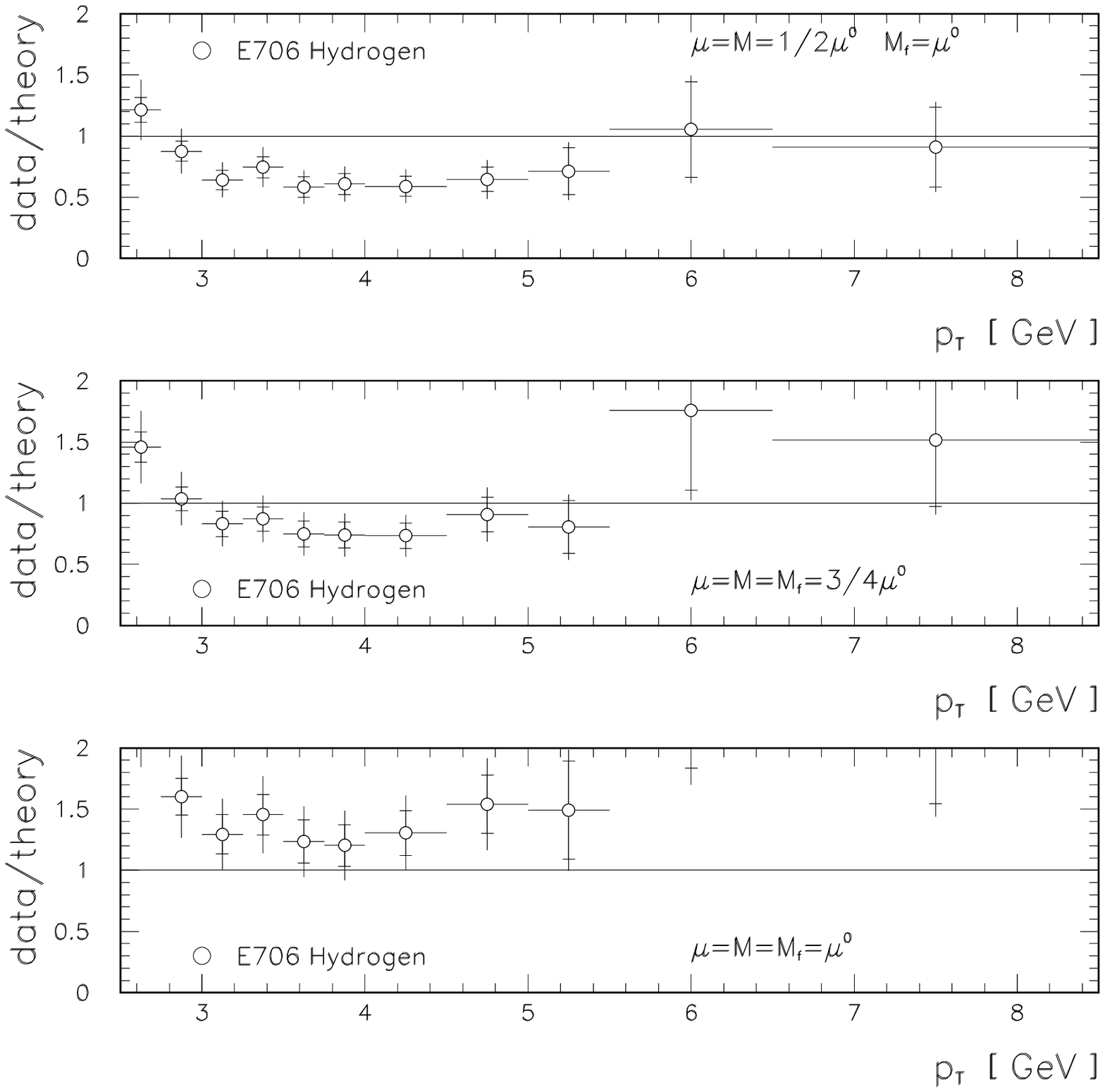}
\end{center}
\caption{\small Scale dependence of the NLO prediction for the $p_{T}$ 
distribution in 800 GeV proton collisions on H target, $\mu^0 =(p_{T1} + p_{T2})/2$.} 
\label{fig5}
\end{figure}

\begin{figure}[htbp]
\begin{center}
\includegraphics[width=0.8\linewidth,height=12 cm]{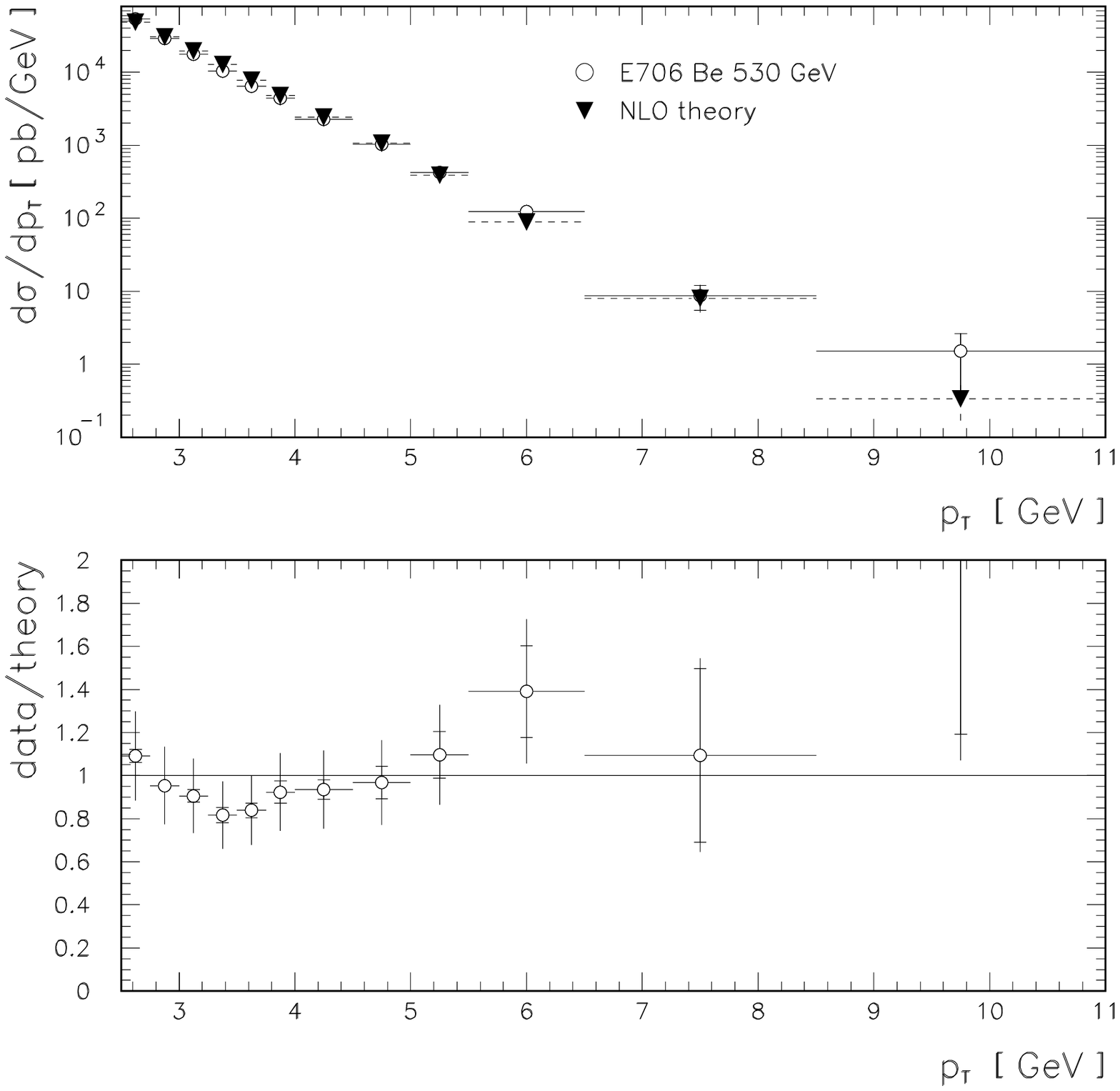}
\end{center}
\caption{\small Dipion differential cross section $d\sigma/dp_T$ vs. $p_T$, 
 the transverse energy of each pion, in $p-Be$ collisions with a beam energy 
 of $530$ GeV.
 Data points with statistical and systematic errors in quadrature 
 are from the E706 collaboration ~\protect\cite{e706}.
 The NLO prediction with scales $M = \mu = M_{f} = 3/8 \, (p_{T1} + p_{T2})$
 is shown as triangles.
 } 
\label{fig1}
\end{figure}

\noindent  
The NLO QCD predictions presented here are provided by the computer code {\em
DIPHOX} presented in the previous section. Both the  theoretical results and
the data are expressed `per effective nucleon' in the Beryllium nucleus (Z=4,
A=9), this effective nucleon being defined by the combination `4/9 proton + 5/9
neutron'. The theoretical calculations use the CTEQ5M set \cite{cteq5} of
parton distribution functions (pdf) for the proton pdf's, and assume that the
neutron pdf's are simply deduced from the ones of the proton by isospin
symmetry. The KKP set \cite{kkp} is used for the fragmentation functions into a pion. Our
theoretical results are given for a common scale choice for the renormalisation
scale $\mu$, initial state factorisation scale $M$ and final state
fragmentation scale $M_{f}$ equal to $\mu = M = M_{f} = (3/4)\mu^0$ where
$\mu^0 =(P_{T1} + P_{T2})/2$ is the average value of the $p_{T}$'s of the
two pions. At fixed target energies, the scale uncertainties remain pretty
large in NLO calculations. 
We made a rapid investigation\footnote{A comprehensive procedure would be, as
in one particle inclusive production, to perform a detailed study of these
dependences with respect to $\mu$, $M$, $M_{f}$ separately. This would help, at
least, in quantifying the scale uncertainties more quantitatively. This would
also possibly determine a region of local stability with select a set of
optimal scales, invoking the Principle of Minimal sensitivity \cite{pms,abfs}.
At fixed target energies, such an optimal scale choice can be found in the case
of single pion inclusive case. However, the surrounding region of local
stability is a narrow area sitting on a sharp ridge \cite{afgkw}, an
unsatisfactory situation. Furthermore, this type of study is quite computer
time consuming already in the single particle inclusive case. Its cost becomes
prohibitive in the case of correlations in dipion production. Therefore we did
not perform such a complete study.} of the scale dependence of the predictions
around the  above choice, using also the two other followings sets of scales:
$\mu = M = (1/2)\mu^0$, $M_{f} = \mu^0$; and $\mu = M = M_{f} =
\mu^0$. The choice $\mu = M = M_{f} = (3/4)\mu^0$ turns out to
accommodate to the data better in the case of the $p_{T}$ distribution of each
pion,  as can be seen in Fig.~\ref{fig1}, whereas the
higher (lower\footnote{One cannot go down to too low scales. Another choice of
scale commonly used would be $\mu = M = M_{f} =  (p_{T1} + p_{T2})/4$ (half 
the average value of the $p_{T}$'s of the two pions). However the lowest
factorisation scale $M_0$ at which the KKP set starts is $M_{0} = \sqrt{2}$
GeV. With the selection cut $p_{T \, min} = 2.5$ GeV on each pion, this choice
would lead to a scale that could fall below $M_{0}$. Technically, a backward
evolution, or, more drastically, the freezing of the pdf's below $M_{0}$ can be
enforced. However the use of such questionable remedies would induce extra
sources of theoretical uncertainties.}) scale choice undershoots (overshoots)
the data, cf. Fig.~\ref{fig5}. In addition, we made comparisons with E706 data
in $p p$ collisions with an energy beam of 800 GeV ($\sqrt{S} \simeq 38.76$
GeV), using the same scale choice and reaching similar conclusions. This
somewhat legitimates the choice of scales adopted. In this section we display
the comparisons between the data and our NLO estimates for the various
observables measured by the E706 experiment. A corresponding phenomenological
analysis will be  presented in the next section. We see that these observables
split qualitatively into two groups, according to the degree of agreement or
disagreement with the experimental spectra.  

\subsection{Well described observables}\label{good-guys}

\begin{figure}[htbp]
\begin{center}
\includegraphics[width=0.8\linewidth,height=12 cm]{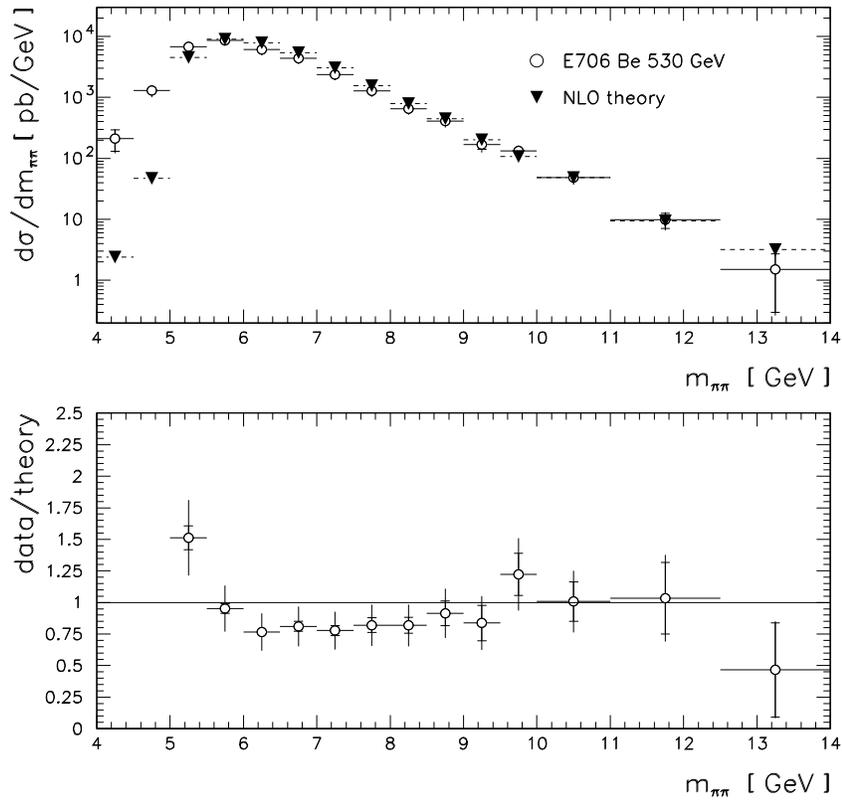}
\end{center}
\caption{\small Dipion differential cross section $d\sigma/dm_{\pi \pi}$ vs. 
 $m_{\pi \pi}$, the invariant mass of the pair, in $p-Be$ collisions 
 with a beam energy of $530$ GeV.
 Data points with statistical and systematic errors in quadrature 
 are from the E706 collaboration ~\protect\cite{e706}.
 The NLO prediction with scales $M = \mu = M_{f} = 3/8 \, (p_{T1} + p_{T2})$
 is shown as triangles. With the chosen ordinate scale in the bottom histogram, the first two points fall outside the plot.
 } 
\label{fig2}
\end{figure}

\begin{figure}[htbp]
\begin{center}
\includegraphics[width=0.8\linewidth,height=12 cm]{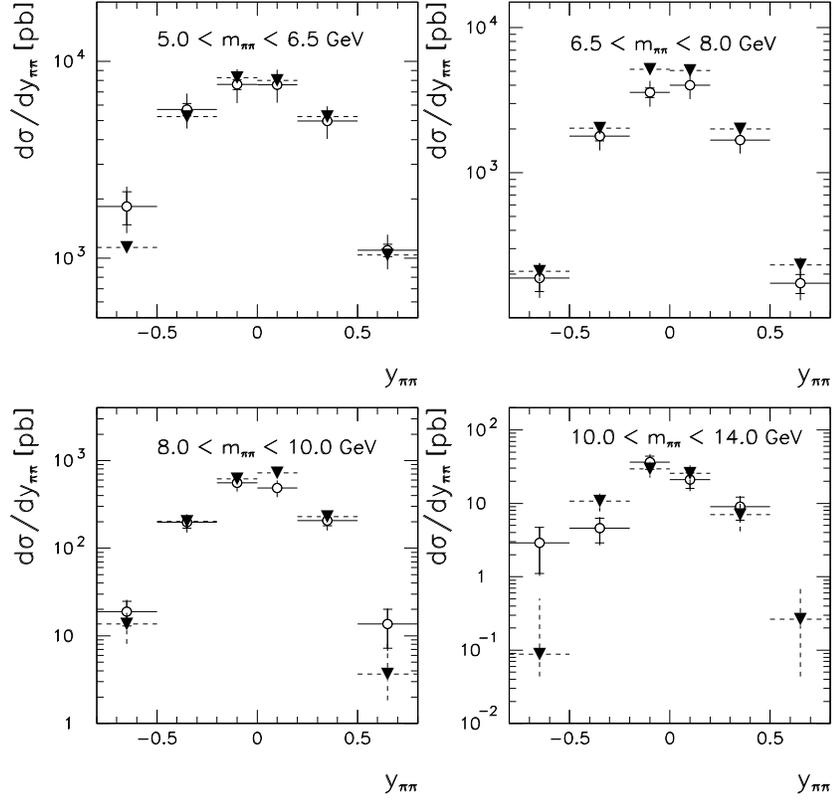}
\end{center}
\caption{\small Dipion differential cross section 
 $d\sigma/dy_{\pi \pi}$ vs. $y_{\pi \pi}$, the rapidity of the pair 
 in $p-Be$ collisions with a beam energy of $530$ GeV. 
 Here $d\sigma/dy_{\pi \pi}$ stands for
 $\int dm_{\pi \pi} (d\sigma/dy_{\pi \pi}dm_{\pi \pi})$
 integrated over the following invariant mass slices: 
  $5.0 < m_{\pi \pi} <  6.5$ GeV (top left), 
  $6.5 < m_{\pi \pi} <  6.5$ GeV (top right),
  $8.0 < m_{\pi \pi} < 10.0$ GeV (bottom left) and 
 $10.0 < m_{\pi \pi} < 14.0$ GeV (bottom right).
 Data points with statistical and systematic errors in quadrature 
 are from the E706 collaboration ~\protect\cite{e706}.
 The NLO prediction with scales $M = \mu = M_{f} = 3/8 \, (p_{T1} + p_{T2})$
 is shown as triangles.
 } 
\label{fig3}
\end{figure}

For the given scale choice, the agreement is reasonably good in shape, as well
as in magnitude, for the distribution of transverse momentum  of each pion in
pairs\footnote{According to this definition each event  enters twice, i.e.:
\[ 
\frac{d\sigma}{dp_{T}} =  
\frac{d\sigma}{dp_{T1}} + \frac{d\sigma}{dp_{T2}} 
\]} $d\sigma/dp_{T}$ (Fig.~\ref{fig1}), for the invariant mass distribution of 
pairs $d\sigma/dm_{\pi \pi}$ (Fig.~\ref{fig2}) - except in the very first bins 
at low invariant mass -, for the rapidity distribution of the pairs $d \sigma/d
y_{\pi \pi}$ (Fig.~\ref{fig3}), and for the  distribution in  $\cos
\theta^{*}$ (Fig.~\ref{fig4bis}). The variable $\cos \theta^{*}$ is defined
by\footnote{This variable has been initially introduced in the literature in the case
of $2 \to 2$ processes, where $\theta^{*}$ is the angle, defined in the center
of mass system of the pair, between the  direction of the beam and the
direction of the pions, emitted back-to-back in this frame. This definition is
convenient at lowest order (LO). 
It is however difficult to extend it in a simple and unambiguous way to 
processes with more than two particles in the final state, as it is
the case in calculation beyond LO in particular. In the case
of a   $2 \to 2$ process, $\cos \theta^{*}$ turns out to be equal to  
$\tanh y^{*}$. In addition to provide an alternative definition which 
is manifestly invariant under longitudinal boosts, this identity allows 
to extend the definition of $\cos \theta^{*}$  to processes with more 
than two particles in the final state. This is the definition used by 
the E706 experiment\cite{e706}, and the one which we adopt here.}  
$\cos \theta^{*} = \tanh y^{*}$ where $y^{*} = (y_{1} - y_{2})/2$  is 
half the difference of the rapidities of the two pions in a pair. 

\vspace{0.2cm}
 
\noindent 
The reasonable agreement observed was expected for these observables over the 
whole measured spectrum. Let us however focus in particular on the distribution
$d \sigma/ d p_{T}$. The remarkable fact is that the agreement is fairly good
over the totality of the measured spectrum. This may be contrasted with the
situation in the case of one-particle inclusive production, measured also by
the same experiment, and which was confronted to a NLO calculation in
Ref.\cite{afgkw}. Let us first briefly remind the results which were obtained
in Ref.\cite{afgkw}. The calculation used the BKK set of parton to pion
fragmentation functions \cite{bkk}. The scales, selected by the application of
the Principle of Minimal Sensitivity \cite{pms,abfs}, were close to $p_{T}/3$.
This set of scale somehow maximized  the theoretical prediction, and turned out
to minimize the discrepancy between the latter and the data. Above $p_{T} \geq$
5.5 GeV, no large difference between the shapes of the NLO calculation and the
data was found. On the other hand, the NLO calculation undershot the data by
about $50 \%$. The discrepancy between the NLO result and the data was found
larger with other, especially larger scales (a factor 1.8 for the scale choice
$p_{T}/2$). Furthermore, in the lower range of the $p_{T}$ spectrum below 5.5
GeV, the slope of the data is increasingly sharper than the one of the NLO
post-diction as the $p_{T}$ decreases. On the contrary in the case of the
$p_{T}$ distribution of each pion in pair production, the agreement is
reasonably good in shape, and also in magnitude with the intermediate scale
choice $\mu = M = M_{f} = (3/8)(P_{T1} + P_{T2})$ used. One notices that the
above scale choice involves higher values (by nearly a factor 2) than the
optimal scales in the case of single inclusive production. The agreement in
shape is rather robust with respect to variations in the arbitrary scales
$\mu$, $M$ and $M_{f}$, at least within the three common values which we
scanned, cf. Fig.~\ref{fig5}. Furthermore, the agreement in the case of pair
production holds even down to fairly low values of $p_{T}$ close to $p_{T \,
min} = 2.5$ GeV above which the events were experimentally selected, in sharp
contrast with the situation in the single inclusive case. This fact may appear
somewhat surprising, and even counterintuitive because one may think that the
agreement should be better, the more inclusive the observable. This puzzle will
be examined in some details in the next section.

\subsection{Poorly described observables}\label{bad-guys}

In contrary, the NLO approximation is completely inadequate to account for the
distribution of azimuthal angle difference between the  two pions,
$d\sigma/d\phi_{\pi \pi}$ (Fig.~\ref{fig6}), and for the distribution $d
\sigma/dp_{T \, out}$ (Fig.~\ref{fig7}). The variable $p_{T \, out}$ is defined
as follows. The direction of the beam axis and the transverse momentum of  one
of the pions define a plane; the component of the $p_{T}$ of the  other pion
which is orthogonal to this plane defines the $p_{T \, out}$  of this
pion\footnote{In other words, the $p_{T \, out}$ of pion $j=1,2$ is equal to
$p_{Tj} \sin \phi_{\pi \pi}$. Notice that each  event enters twice in this
distribution.}. The disagreement is also sharp for the transverse momentum
distribution of pairs $d\sigma/dq_{T}$ (Fig.~\ref{fig8}). The transverse
momentum balance distribution $d\sigma/dZ$ (Fig.~\ref{fig9}) is not adequately 
described by the NLO approximation either, particularly in the region $Z < 1$. 
The $p_{T}$ balance variable $Z$ is defined for the pion
$j$ by\footnote{In this observable as well each event enters twice.} $Z = -
(\vec{p}_{T1} \cdot \vec{p}_{T2})/|p_{Tj}|^{2}$. The shapes of these
observables, especially of $\phi_{\pi \pi}$ and $p_{T \, out}$ distributions
are notoriously infrared sensitive respectively when $\phi_{\pi \pi} \to \pi$
and $p_{T \, out} \to 0$. Therefore it is not surprising that the NLO
calculations do not describe these regions adequately. The noticeable fact is
that even the tails of these distributions are unmatched. The NLO estimates
fall much too rapidly compared with the data, the former being more than one
order of magnitude below the latter. These features will be examined in detail
in the next section. Last, in Fig.~\ref{fig2} we notice also the poor account of the very first
bins at low invariant mass in the invariant mass distribution by the NLO
calculation.

\section{A classification of dipion observables}\label{phenom}

The observables of inclusive pion pair production can be theoretically
classified into two main categories, according to their respective infrared
sensitivity. This theoretical classification matches with the phenomenological
one above, according to the `degree of agreement (or disagreement)' between the
NLO predictions provided by {\em DIPHOX} for inclusive neutral pion pair
production and the recent high statistics data of the fixed target E706
experiment at Fermilab. 

\vspace{0.2cm}

\noindent       
The first category which we will consider gathers the observables which are
infrared insensitive. By this we mean that the contribution associated with the
multiple soft gluon emission is not logarithmically enhanced order by order in
perturbation theory. Typical representatives of this class of observables are:
the distribution $d\sigma/d p_{T}$; and $d \sigma/dm_{\pi \pi}$, at least when
the invariant mass $m_{\pi \pi}$ is larger than  $2 \, p_{T \, min}$
(the reason for this restriction will be examined later). Other examples are:
the distribution $d\sigma/d \cos \theta^{*}$; and the rapidity distribution of
the pairs, $d \sigma/dY$.

\vspace{0.2cm}

\noindent   
The opposite class gathers the infrared sensitive distributions. The shapes of
these distributions are well known to be strongly influenced by multiple soft
gluon emission in the vicinity of some `critical point'. Depending on the
observable considered, this critical point may lie at the elastic boundary of
the spectrum; or it can stand inside the physical domain. Order by order in
perturbation theory the infrared sensitivity is signed by the appearance of
large logarithms of Sudakov type which become infinite at the critical point.
Among these distributions, one may distinguish between two types. A first
family is made of the distributions for which the perturbative calculation is
singular at the critical point order by order in perturbation theory. In the
present case of pion pair production, typical representatives are
$d\sigma/d\phi_{\pi \pi}$, $d \sigma/d p_{T \, out}$. Any fixed order
calculation is certainly inadequate to estimate this type of observables in the
vicinity of the critical point, and, whenever possible, a Sudakov type
resummation has to be carried out to restore a predictive power of the
perturbative QCD calculation. As will be commented in subsect.
\ref{ir-sensitive}, the colour structure of the hard subprocess in
hadroproduction of pairs of hadrons is more complicated than in the celebrated
case of the Drell-Yan process. This makes a soft gluon summation beyond the
leading logarithmic accuracy quite sophisticated in this case. We do not
provide a study of these observables based on a resummed calculation. A second
family gathers observables which are less infrared sensitive than the previous
ones, in the sense that their fixed order estimates do not diverge at the
critical point; therefore the inadequacy of fixed order calculation to account
for such observables is more difficult to foresee. Among these observables, let
us mention the distribution of transverse momentum of pairs, $d\sigma/d q_{T}$,
and the  distribution $d \sigma/ d Z$ where the variable $Z$ has been defined
in subsect. \ref{bad-guys}. They would belong to the first family, if the
fragmenting partons in the final state were, instead, observed particles - as,
e.g. it would be the case if they were prompt photons. In the collinear
fragmentation model used here, the screening of the infrareed sensitivity stems from the convolution with fragmentation functions. this provides extra
integrations over longitudinal fragmentation variables. 

\vspace{0.2cm}

\noindent 
Computing the NLO truncation for infrared sensitive observables is more than an
academic exercise. Indeed, whereas the vicinity of the critical point is
infrared sensitive, the tail of the distribution may be expected safer for
fixed order perturbative calculations. Yet the infrared sensitivity of these
observables connects the perturbative and non perturbative regimes. At fixed
target energies, depending on the observable, on the selection cuts, etc., the
non perturbative effects may happen to be not confined to the vicinity of the
critical point, but instead can spread over a large fraction of the physical
spectrum, if not the whole. A confrontation of the NLO truncation to the
measured spectrum can give a hint of the depth to which infrared, in particular
non perturbative, effects penetrate a distribution away from the infrared critical point. This is what
we have in mind while performing this confrontation here.  Another interest
concerns observables which are infrared insensitive apart from a few bins
where, for example, kinematic cuts make them locally semi inclusive. The
discrepancy between the NLO prediction and the measured spectrum in these bins
can be  related to some infrared sensitive distribution inadequately accounted
for by fixed order truncation. This is typically the case with the lowest bins
of the invariant mass distribution, where the discrepancy between NLO and data
can be understood in the light of $d\sigma/d\phi_{\pi \pi}$, as will be
explained below.

\vspace{0.2cm} 

\noindent  
Let us add a comment about the classification which we present here, especially
regarding the infrared sensitive observables. Phenomenological models are
sometimes used for `effective' $k_{T}$ dependent distribution - and
fragmentation - functions. For example, a popular procedure
\cite{baier-engels-peterson,owens-rmp} revisited recently \cite{apanasevitch}
assumes a $x$ independent, gaussian $k_{T}$  distribution functions for on
shell partons. In such a model, the behaviour of distributions such as 
$d\sigma/d \phi_{\pi \pi}$ and $d \sigma/d p_{T \, out}$ at respectively
$\phi_{\pi \pi} \to \pi$ and $p_{T \, out} \to 0$ are smeared by the 
convolutions with the $k_{T}$ distributions. Therefore, according to such a
model, all distributions would be affected by $k_T$ effect, i. e. the first
family would be empty. In order to avoid any confusion in the terminology used,
let us emphasize the following point. The models of $k_{T}$  smearing mentioned
above {\it depart} from the property of collinear factorisation, and they do
not account for the higher orders corrections to the hard subprocess in
perturbation theory in a consistent way. On the contrary, the classification
which we present here follows closely the property of collinear factorisation
which holds in perturbative QCD.

\subsection{Infrared insensitive observables}\label{ir-insensitive}

\subsubsection*{Transverse momentum: one-particle inclusive vs. inclusive pair
production}

Let us analyse the difference between the $p_{T}$ distribution of each pion in
the present case of dipion production, and the $p_{T}$ distribution in one
particle inclusive production studied in Ref.~\cite{afgkw}. We have to
disentangle two issues. The one concerns the difference in the ingredients
used in Ref.~\cite{afgkw} vs. the present work. It will affect the whole
spectrum, and it is especially relevant in the larger $p_{T}$ range. The second
concerns the physics involved, and will particularly affect the lower range of
the spectrum. We consider both of them successively.

\vspace{0.2cm}

\noindent
First, the sets of fragmentation functions used in either case are different. 
This difference turns out to be quite significant in the case of single pion
inclusive production. It affects the whole spectrum, and it matters especially
in the higher range of the $p_{T}$ spectrum, $p_{T} \geq 5$ GeV, where we
expect perturbative QCD to apply. Indeed, as seen in Fig.~\ref{fig10}, 
the $u$ and $d$ fragmentation
functions of the KKP set are  larger than the corresponding ones of the BKK
set: more than 50 \% above $z = 0.7$, and the ratios increase as $z$ increases
towards 1. The difference (in the same direction) is even  larger for the gluon
fragmentation: more than a factor 2. This is possible because the sets of
fragmentation functions are extracted from data coming from $e^{+}e^{-}$
annihilation into hadrons which constrain the fragmentation functions
essentially in the range $0.1 \leq z \leq 0.7$ and because, unlike pdf's, the
fragmentation functions are not tightly constrained by a momentum sum
rule\footnote{In the framework of independent fragmentation functions, there
exists indeed a sum rule:   
\[ 
\sum_{H} \int_{0}^{1} dz \, z\, D_{H/a}(z,M_{f}^{2}) = 1   
\]   
The latter means that a parton of given species $a$ fragments into any
hadron species $H$ with probability 1. It is however much less
constraining in practice than the one holding for pdf's which relates the
pdf's of all the species for a given incoming hadron.}. 

\vspace{0.2cm} 

\noindent
It is worth noticing that for $p_{T}$ ranging from 4 to 10 GeV, the average
value $<z>$ of the fragmentation variable $z$ in the one particle inclusive
case is  $<z> \sim 0.75 - 0.90$. Therefore the theoretical estimates for single
inclusive production rely on extrapolations of the fragmentation functions
outside the region where they are actually constrained by the data from which
they are extracted. Consequently, as emphasized in Ref.~\cite{afgkw}, the NLO
predictions are plagued by a rather large uncertainty.  The effect of the
replacement of the BKK set by the KKP one is shown in Fig.~\ref{fig11}. The
discrepancy in normalisation between  the theoretical prediction and the data
is strongly reduced, for the scale choice $p_{T}/3$. 

\vspace{0.2cm}

\noindent
Comparatively, the theoretical calculation of inclusive pion pair production
is less affected by the uncertainties on fragmentation functions at large $z$.
The reason is that the values of the fragmentation variables $z_{1}$, $z_{2}$
which dominate the production of pion pairs extends in a region of smaller
values than the $<z>$ in the single inclusive case, $z_{1,2}$ typically $\sim
0.5$, where the $e^{+}e^{-}$ data constrain the fits of the fragmentation
functions quite well. This feature is illustrated by Fig.~\ref{fig12}. This can be traced back to the fact that the region in $z$, which
dominates, results from a competition between the cost in energy momentum
involved in the partonic subprocess -- the pdf's tend to favour the smallest
possible $x$ fractions for incoming energy-momenta and therefore larger values
of $z$ -- and the price to pay for a fragmentation near the end point $z = 1$ --
the fragmentation stage favours small values of $z$. The production of a pair of
two hard particles involves two fragmentation functions instead of one as in
the single inclusive case, which enforces the role of fragmentation in the
competition. The situation will be very different regarding the reducible
background to Higgs boson search at LHC \cite{newcitation}, because isolation requirements will
select the large $z$ region. This issue will be addressed in detail in a
forthcoming article \cite{bgpw2}.

\vspace{0.2cm}

\noindent 
The difference in the dominant ranges of $z$ in one particle inclusive vs. pair
production sheds also some light on the apparent difference in scales which
seem to be preferred by the data in either case, the optimal scale choice $\sim
p_{T}/3$ in the one particle is apparently smaller than the choice $(3/8) \,
(p_{T1}+p_{T2})$ used in pair production. Let us consider the inclusive
production of a single pion in the region where the E706 data agree reasonably
well with the NLO calculation using the KKP set, say, near $p_{T} = 6$ GeV.
Such pions probe a partonic regime  with a transverse momentum $\sim p_{T}/\!<\!z\!> \,
\sim 7$ GeV, and the optimal scale is $\sim 2$ GeV. A pair of pions, dominantly
back to back, with individual $p_{T} \sim 3$ GeV probes a parton regime with
parton transverse momenta $\sim p_{T}/\!<\!z_{1,2}\!>$ comparable with the previous
one; we notice also that the scale choice used in  this case is $9/4 \sim 2$
GeV. Consequently, the same partonic kinematic regime is probed in these two
cases, and are found to be consistently in reasonable agreement with the same 
theoretical framework. We therefore conclude that the situation is actually
comparable between the two cases in this $p_{T}$ range. Of course this rapid
investigation considers only the comparison with the data of E706. To obtain a
comprehensive understanding, one needs to perform a complete study revisiting
single inclusive production at fixed target energy involving other fixed
targets experiments. 

\vspace{0.2cm}

\noindent 
Let us now examine the difference in the behaviours of the $p_{T}$ spectra of
single inclusive production vs. each pion in pair production in the lower range
of the $p_{T}$ spectrum. The NLO calculations performed in the QCD improved
parton model according to the property of collinear factorisation neglect non
perturbative, power corrections $\sim {\cal O} \left( 1 \,\mbox{GeV}/p_{T})^{n} \right)$ where $n$ depends on the observable considered.
Therefore this `leading twist' approximation is valid when $p_{T} \gg {\cal
O}(1 \, \mbox{GeV})$. On the other hand, NLO calculations are not expected to
predict correctly inclusive $p_{T}$ spectra when $p_{T}$ becomes too small, and
distortions of the perturbative QCD predictions are expected up to $p_{T} \sim$
a few GeV. One may instead wonder why the NLO calculation and the data agree
well on the slope of the distribution of each pion's $p_{T}$ for pair
production down to $p_{T} \sim p_{T \, min} =  2.5$ GeV.

\vspace{0.2cm}

\noindent 
Among the neglected effects, the ones induced by finite transverse momenta
($k_{T}$) of incoming partons are often assumed to be the most important. This 
concerns in particular the understanding of the $p_{T}$ distribution at low
$p_{T}$  in one particle inclusive production due to the so-called `trigger
bias' effect, tentatively visualized in Fig.~\ref{thomas-circle1}.
A current practice models the latter by assuming an effective $k_{T}$
distribution function\footnote{For example, the popular gaussian model
mentioned above.} for the incoming partons, still considered on shell, and 
convolutes this with a LO calculation. If one notes $g(k_{T})$ this $k_{T}$ 
distribution, and assuming that it depends only on $|\vec{k}_{T}|$, 
one has schematically
\begin{eqnarray}
\frac{d \sigma}{d p_{T}} & \propto & 
\int d^{2}k_{T} \, g(k_{T}) \,  \frac{1}{| \vec{p}_{T} - \vec{k}_{T}|^{\alpha}} \label{kt}
\nonumber\\
& \sim & \frac{1}{p_{T}^{\alpha}} \left( 1 + 
\frac{\alpha^{2}}{2} \frac{<\! k_{T}^{2} \!>}{p_{T}^{2}} + 
{\cal O} \left( \frac{<\!k_{T}^{4}\!>}{p_{T}^{4}} \right) \right)
\nonumber
\end{eqnarray}
\noindent 
This means that the contribution from subprocesses initiated by partons with
incoming transverse momentum pointing in the direction of the measured pion is
enhanced, and this leads to an increase of the cross section in average,
especially at low $p_{T}$.  Let us however stress that this model is not
consistent\footnote{In addition it requires the use of several ad hoc extra
parameters and cuts, in particular to prevent the convolution integral in
eqn.(\ref{kt}) from diverging. The divergence of eqn. (\ref{kt}) when $k_{T} =
p_{T}$ corresponds to a soft partonic subprocess boosted transversally by the
$k_{T}$ kick. The divergence obtained is an artefact of the model. Another
cut-off has to be introduced to prevent the longitudinal momentum of an
incoming parton $a$ and the momentum of its parent hadron $H$ from being in
opposite directions.  This nasty can occur at non zero $k_{T}$ since  \[ p_{a
//} = \frac{1}{2} \left( x_{a} P_{H} -  \frac{k_{T \, a}^{2}}{x_{a} P_{H}}
\right) \] where $x_{a}$ is the light cone momentum fraction of $H$ carried by
$a$. This problem is an artefact of the modelling (an on shell parton with
non zero transverse momentum and a Lorentz invariant momentum fraction). See
\cite{owens-rmp} for a review.} with the property of collinear factorisation,
established in perturbative QCD in the leading twist approximation, for
inclusive processes with hard probes. Attempts of a consistent formulation have
been proposed recently \cite{lai-li,sterman-laenen-vogelsang}. This formalism
is still in an exploratory stage, and no firm and quantitative conclusions can
be drawn yet.

\vspace{0.2cm}

\noindent  
On the other hand, the distribution of each pion's $p_{T}$ in pair production
involves the detection of {\it two} hard particles, dominantly back-to-back in
the transverse plane. Therefore, any $k_{T}$ kick which would favour one pion
in a pair would penalize the other \cite{baier-engels-peterson}. This is
tentatively pictured by the top drawing in Fig.~\ref{thomas-circle2}. In 
particular, in the 
lower range of the
$p_{T}$ spectrum, the $p_{T}$ of the penalized pion may even be pulled down
below the selection cut $p_{T \, min}$ so that the event may be discarded, 
cf. bottom drawing in Fig.~\ref{thomas-circle2}.
Consequently, when the selection cut is symmetric, the mentionned trigger bias is reduced. If the mechanism of $k_{T}$ gives the dominant power
correction in the single inclusive $p_{T}$ spectrum at low $p_{T}$, its lesser
influence on the $p_{T}$ spectrum of each pion in pair production allows
perturbative QCD to extend its range of validity down to fairly low $p_{T}$
values. Let us notice that the absence of trigger bias effects is expected as 
well if, instead of the distribution of each pion's $p_{T}$, but with the same
selection cuts, one would consider the $p_{T}$ distribution of the `leading'
pion (the one with the largest $p_{T}$ in the pair), even though the latter
seemingly displays the $p_{T}$ of only one particle. On the other hand, an
interpolating case would be the distribution of each pion's $p_{T}$ in pair
production when the selection cuts on the two pions are asymmetric. The
behaviour of the distribution of each  pion's $p_{T}$ is expected to exhibit
the same trigger bias effect as the single inclusive case when the $p_{T}$ cut
on the second pion becomes very low.

\subsubsection*{The invariant mass distribution of pairs, 
$d \sigma/dm_{\pi \pi}$}

The {\em DIPHOX} calculation and the E706 data agree reasonably well on this
observable for $m_{\pi \pi} \geq 5.0$ GeV. On the other hand we see a sharp
disagreement on the first two bins in invariant mass. In this respect we notice
that 
\begin{equation}\label{C} 
m_{\pi \pi} = \sqrt{2 p_{T1} \, p_{T2} (\cosh (2y^{*}) - \cos \phi_{\pi \pi})}
\end{equation}
\noindent
$\phi_{\pi \pi}$ being the azimuthal angle between the two pions. At
LO, the $2 \to 2$ kinematics forces the $p_{T}$ of the two pions to be
back-to-back ($\cos \phi_{\pi \pi} = -1 $) and the value 5 GeV, which
corresponds to 2 $p_{T \, min}$, is the minimal value which can be taken by
$m_{\pi \pi}$ at LO. In other words, the first two bins of the theoretical
calculation are filled only by NLO, inelastic contributions, for which
$\phi_{\pi \pi} < \pi$. The sharp disagreement between the NLO calculation and
the data in these two bins is therefore correlated with the situation regarding
the distribution in azimuthal angle difference between the two pions in a pair,
which will be examined in sect. \ref{ir-sensitive}. 

\subsubsection*{The distribution $d \sigma/d \cos \theta^{*}$}

\begin{figure}[htbp]
\begin{center}
\includegraphics[width=0.8\linewidth,height=12 cm]{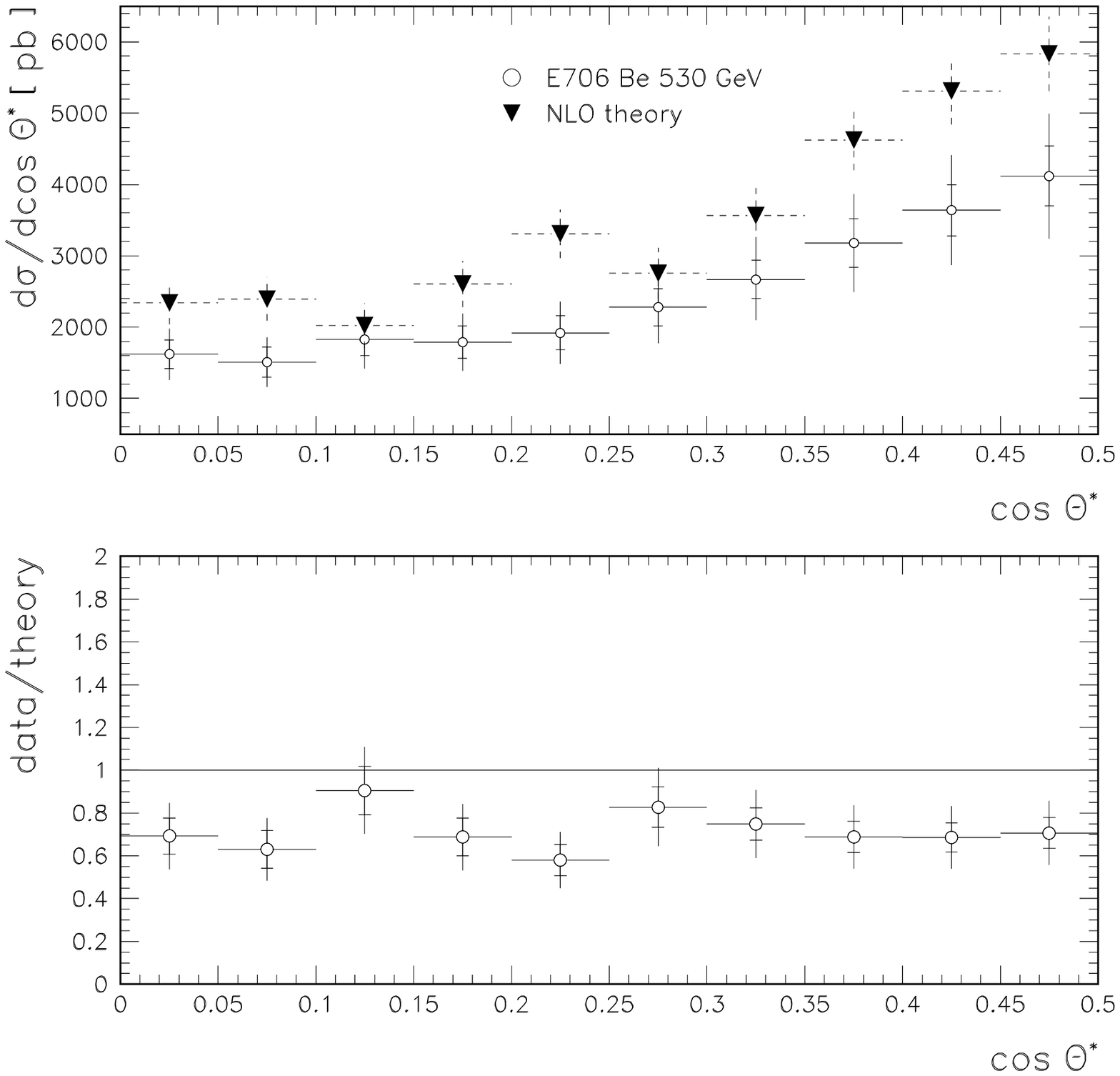}
\end{center}
\caption{\small Dipion differential cross section $d\sigma/d\cos \theta^{*}$ vs. 
 the variable $\cos \theta^{*}$ defined in subsect. \protect\ref{good-guys}, in $p-Be$ 
 collisions with a beam energy of $530$ GeV.
 Data points with statistical and systematic errors in quadrature 
 are from the E706 collaboration ~\protect\cite{e706}.
 The NLO prediction with scales $M = \mu = M_{f} = 3/8 \, (p_{T1} + p_{T2})$
 is shown as triangles.
 } 
\label{fig4bis}
\end{figure}

\begin{figure}[htbp]
\begin{center}
\includegraphics[width=0.8\linewidth,height=12 cm]{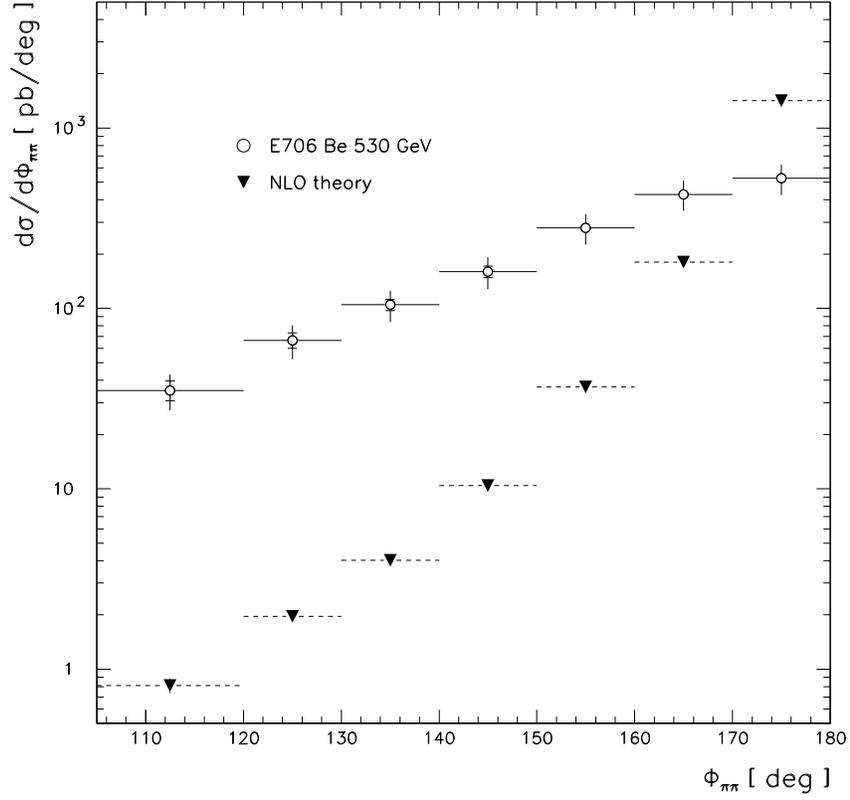}
\end{center}
\caption{\small Dipion differential cross section $d\sigma/d\phi_{\pi \pi}$ vs. 
 $\phi_{\pi \pi}$, the azimuthal angle difference between the two pions in a 
 pair, in $p-Be$ collisions with a beam energy of $530$ GeV.
 Data points with statistical and systematic errors in quadrature 
 are from the E706 collaboration ~\protect\cite{e706}.
 The NLO prediction with scales $M = \mu = M_{f} = 3/8 \, (p_{T1} + p_{T2})$
 is shown as triangles.
 } 
\label{fig6}
\end{figure}

\begin{figure}[htbp]
\begin{center}
\includegraphics[width=0.8\linewidth,height=12 cm]{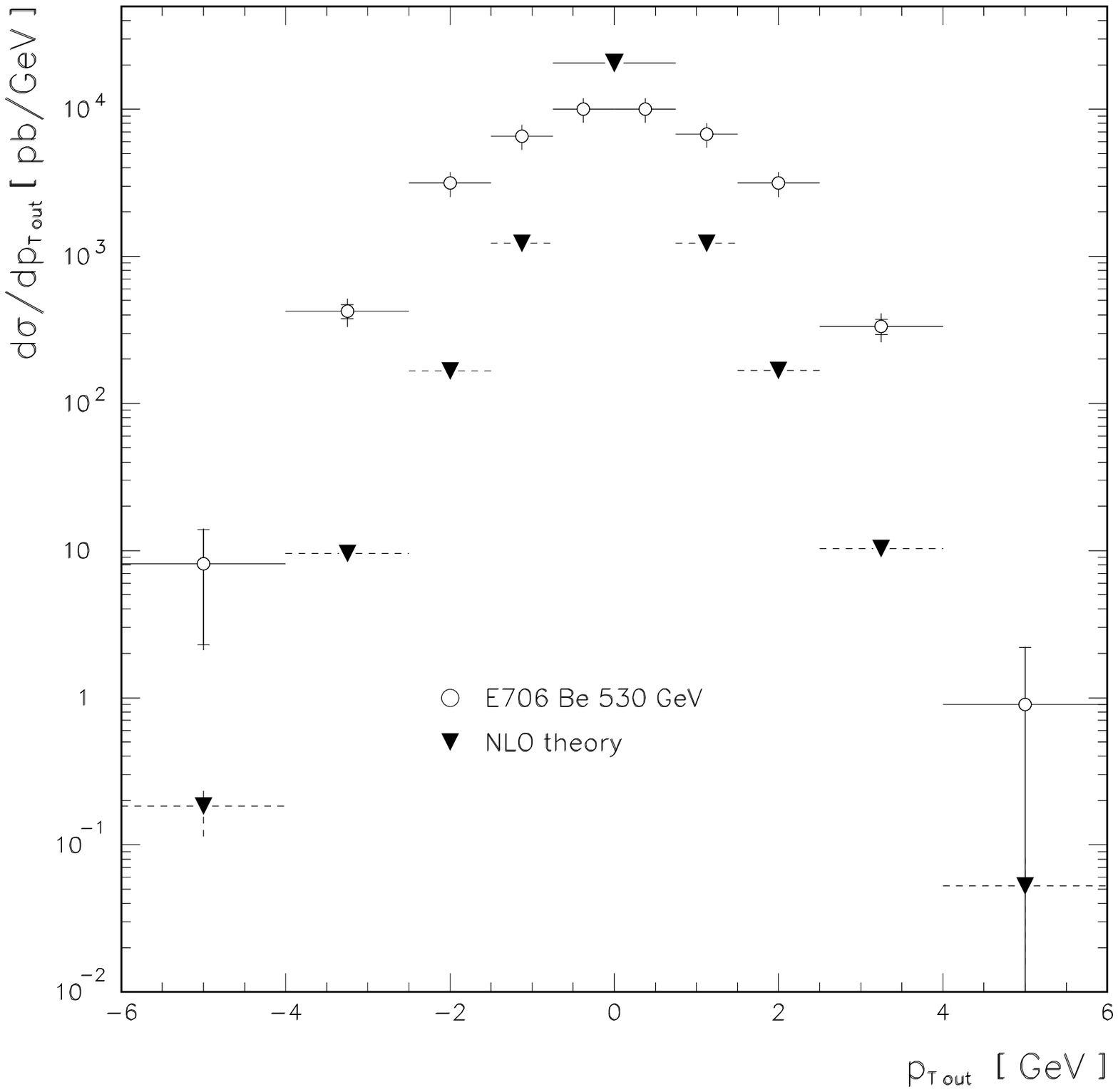}
\end{center}
\caption{\small Dipion differential cross section $d\sigma/dp_{T , out }$ vs. 
 the variable $p_{T , out }$ defined in subsect. \protect\ref{bad-guys}, in $p-Be$ 
 collisions with a beam energy of $530$ GeV.
 Data points with statistical and systematic errors in quadrature 
 are from the E706 collaboration ~\protect\cite{e706}.
 The NLO prediction with scales $M = \mu = M_{f} = 3/8 \, (p_{T1} + p_{T2})$
 is shown as triangles.
 } 
\label{fig7}
\end{figure}

\begin{figure}[htbp]
\begin{center}
\includegraphics[width=0.8\linewidth,height=12 cm]{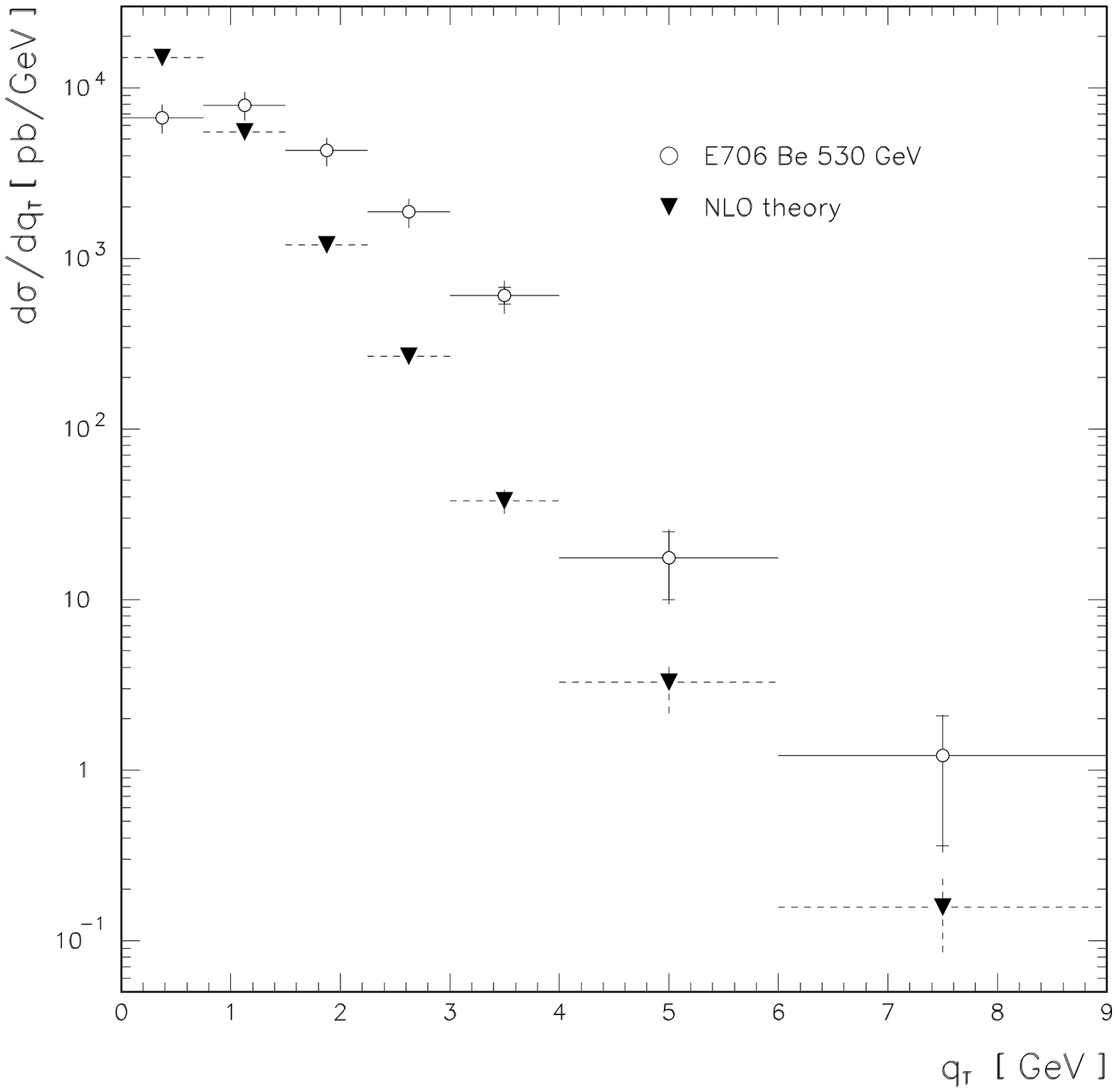}
\end{center}
\caption{\small Dipion differential cross section $d\sigma/dq_{T}$ vs. 
 $q_{T}$, the transverse momentum of a 
 pair, in $p-Be$ collisions with a beam energy of $530$ GeV.
 Data points with statistical and systematic errors in quadrature 
 are from the E706 collaboration ~\protect\cite{e706}.
 The NLO prediction with scales $M = \mu = M_{f} = 3/8 \, (p_{T1} + p_{T2})$
 is shown as triangles.
 } 
\label{fig8}
\end{figure}

\begin{figure}[htbp]
\begin{center}
\includegraphics[width=0.8\linewidth,height=12 cm]{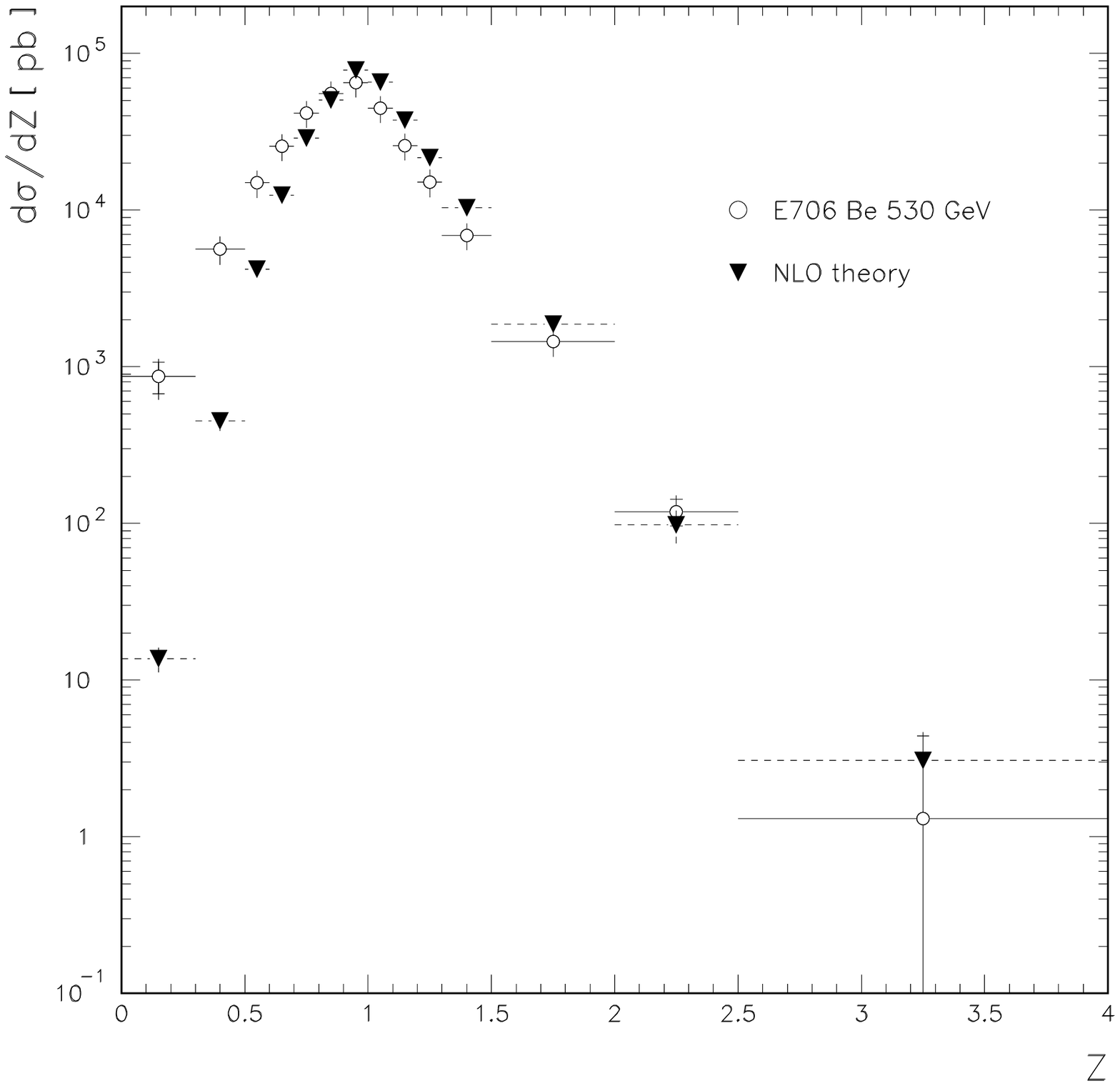}
\end{center}
\caption{\small Dipion differential cross section $d\sigma/dZ$ vs. the variable
 $Z$ defined in subsect. \protect\ref{bad-guys}, in $p-Be$ collisions with a beam 
 energy of $530$ GeV. 
 Data points with statistical and systematic errors in quadrature 
 are from the E706 collaboration ~\protect\cite{e706}.
 The NLO prediction with scales $M = \mu = M_{f} = 3/8 \, (p_{T1} + p_{T2})$
 is shown as triangles.
 } 
\label{fig9}
\end{figure}

\begin{figure}[htbp]
\begin{center}
\includegraphics[width=0.8\linewidth,height=12 cm]{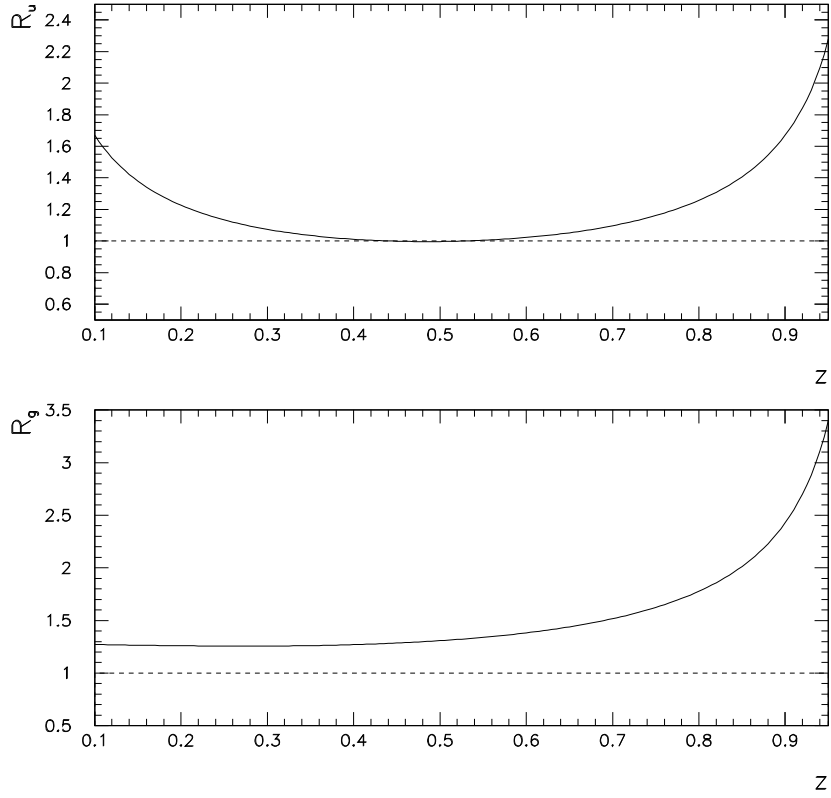}
\end{center}
\caption{\small Ratios 
$R_{f}(z) = D_{\pi/f}^{KKP}(z,M_{f})/D_{\pi/f}^{BKK}(z,M_{f})$
of BKK to KKP fragmentations functions, vs. $z$,  
for various partons flavours $f$, at $M_{f}=$ 3 GeV. Top: $f = u$ quark 
($f = d$ quark equal to $u$). Bottom: $f$ = gluon. 
 } 
\label{fig10}
\end{figure}

\begin{figure}[htbp]
\begin{center}
\includegraphics[width=0.8\linewidth,height=12 cm]{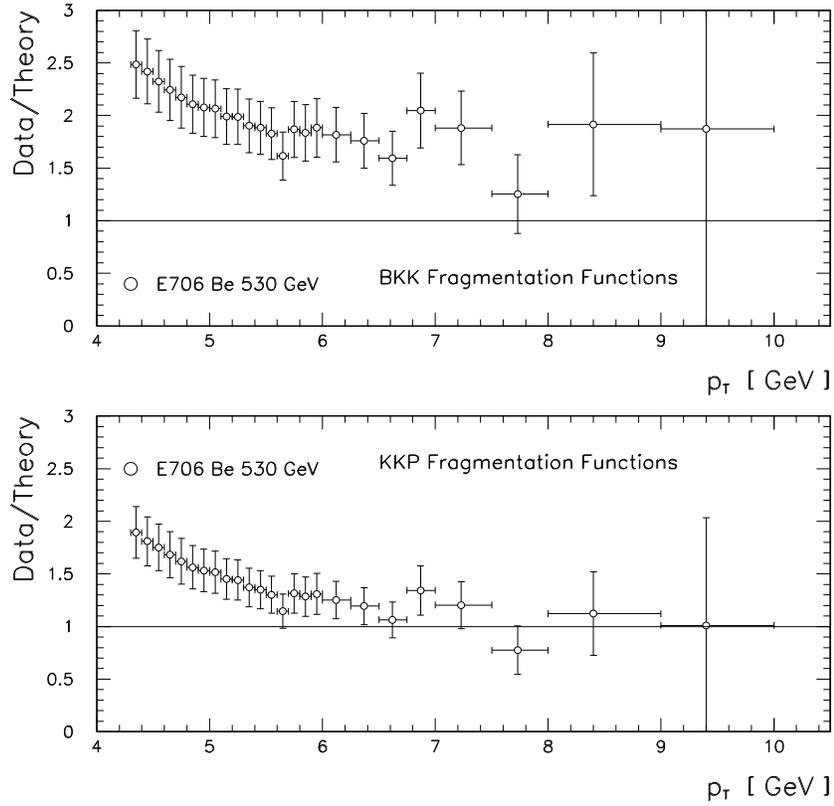}
\end{center}
\caption{\small Collisions of 530 GeV protons on Be
target. Comparisons of E706 data with NLO predictions for the one pion 
inclusive $p_{T}$ spectrum, for BKK and KKP sets of fragmentation functions. 
The scale choice used is $\mu = M = M_{f} = p_{T}/3$. Only data points with
$p_{T} > 4.35$ GeV are kept in order that the fragmentation
scale $M_{f}$  in the NLO predictions is above the starting scale $\sqrt{2}$ 
GeV of the BKK and KKP sets.
 } 
\label{fig11}
\end{figure}

\begin{figure}[htbp]
\begin{center}
\includegraphics[width=0.8\linewidth,height=12 cm]{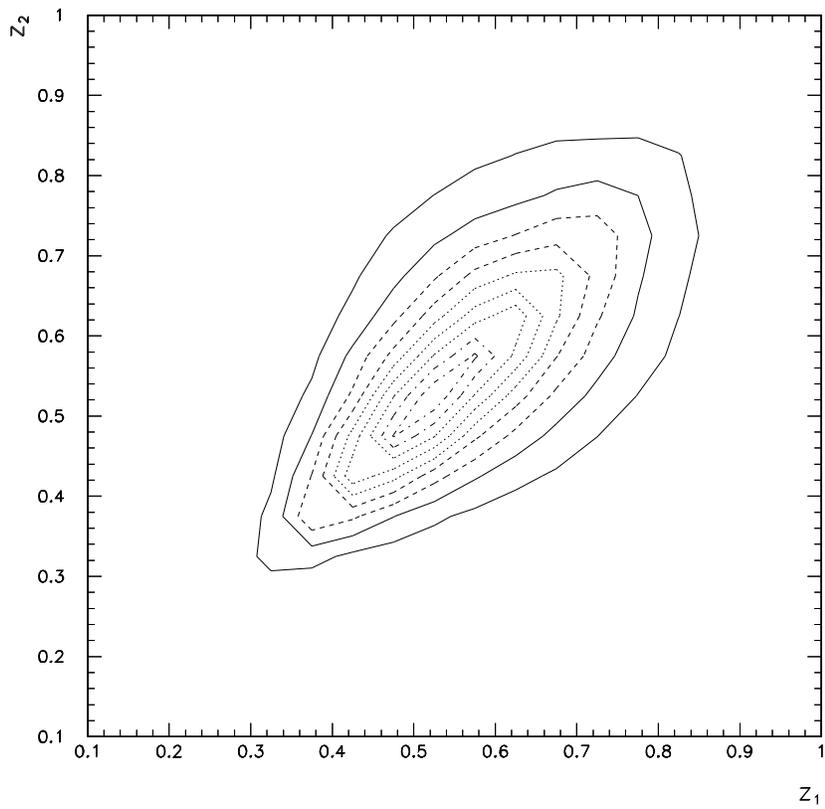}
\end{center}
\caption{\small Contour plot of $A = d\sigma/dz_{1}dz_{2}$ in the plane of the 
longitudinal fragmentation variables $(z_{1},z_{2})$. The iso-$A$ curves are 
oriented along the diagonal $z_{1} = z_{2}$. The most inner iso-$A$ curves, 
centered on the region $z_{1} = z_{2} \simeq 0.5$ correspond to the largest 
values of $A$.
 } 
\label{fig12}
\end{figure}

\begin{figure}[htbp]
\begin{center}
\includegraphics[scale=0.7]{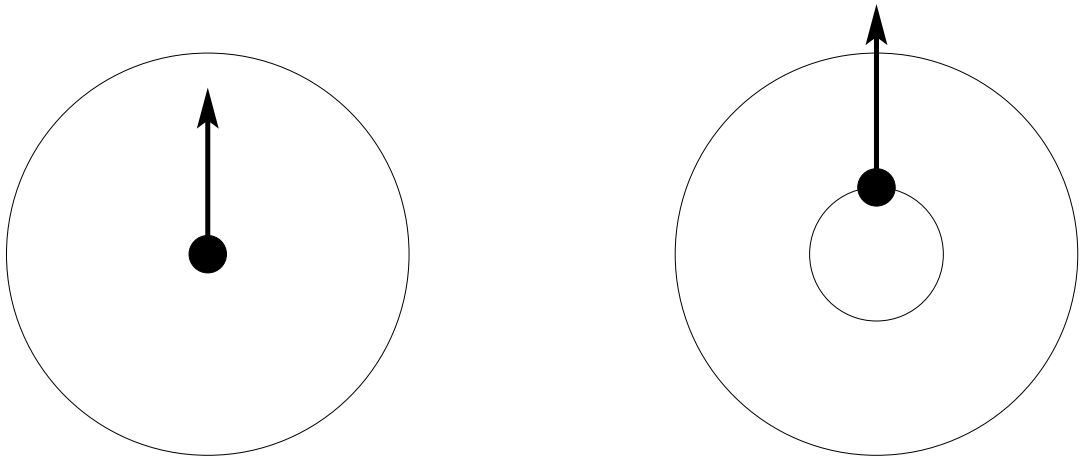}
\end{center}
\caption{\small Illustration of trigger bias effect in one particle inclusive 
 production. The large circle has radius $p_{T \, min}$. The thick arrow stands
 for the transverse momentum of the pion generated by the hard subprocess.
 Left: pion not hard enough and no $k_{T}$ kick: event discarded . The
 radius of the inner circle on the right indicates the magnitude of the 
 $k_{T}$ kick. Right: the pion would not be produced hard enough, but the 
 partonic subprocess is kicked transversally: event accepted.}
\label{thomas-circle1}
\end{figure}

\begin{figure}[htbp]
\begin{center}
\includegraphics[scale=0.7]{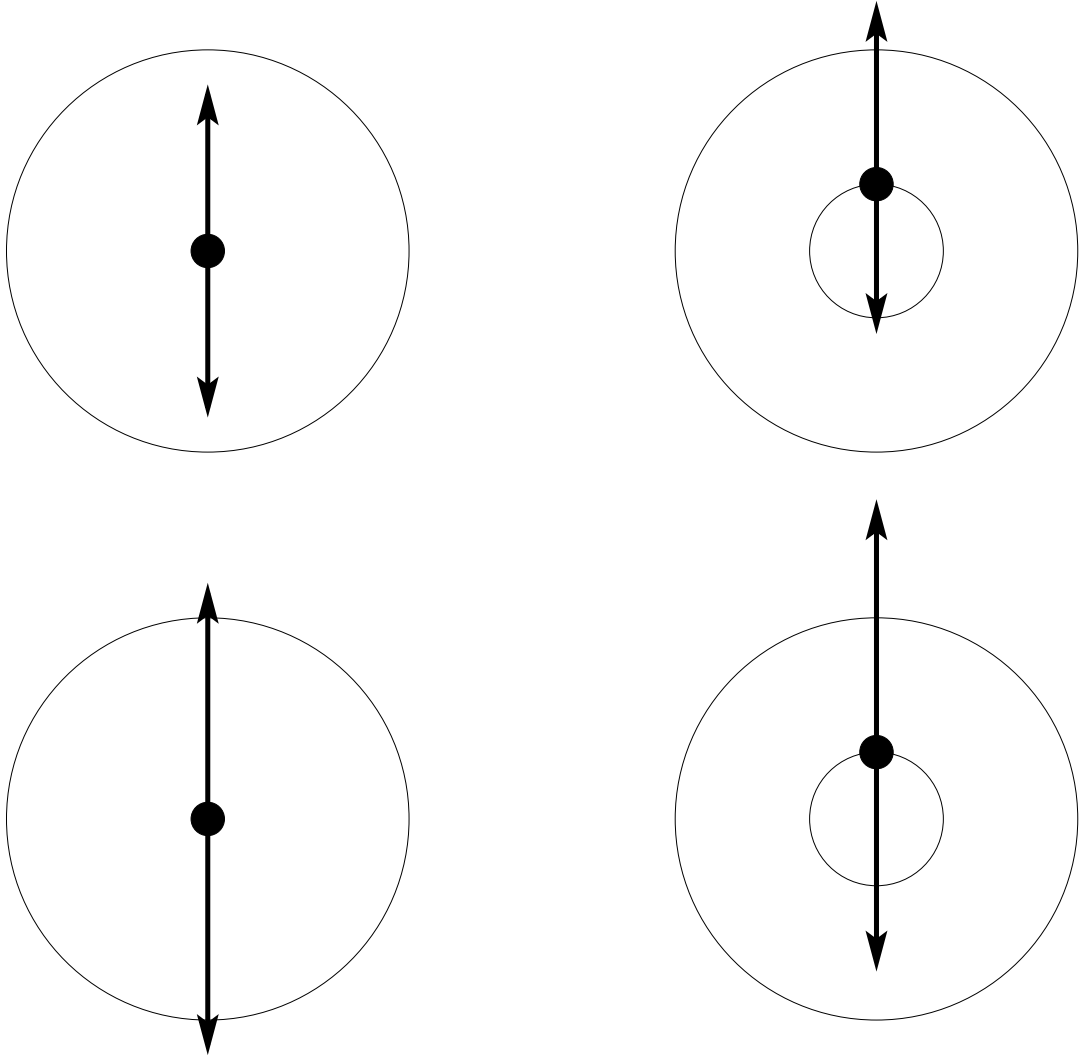}
\end{center}
\caption{\small Illustration of the absence of trigger bias effect in inclusive 
 pair production. The large circle has radius $p_{T \, min}$. The thick arrows 
 stand for the transverse momenta of the pions generated by the hard 
 subprocess. The radius of the inner circle on the right indicates the 
 magnitude of the $k_{T}$ kick. Top: the pions are not produced hard enough by
 the partonic subprocess. Top left: no $k_{T}$ kick: event discarded. 
 Top right: one pion is favoured by $k_{T}$ kick, the other one is penalized: 
 event also discarded. Bottom: the pions would be produced hard enough by the
 subprocess. Bottom left: no $k_{T}$ kick, event accepted. Bottom right: one
 pion is kicked down, even discarded.
 }
\label{thomas-circle2}
\end{figure}

As seen on Fig.~\ref{fig4bis}, the measured distribution  $d\sigma/d\cos
\theta^{*}$ and the corresponding NLO distribution show a fairly good agreement
in shape but an overall discrepancy in normalisation\footnote{A current 
presentation of this observable normalises the measured distribution and the
NLO prediction separately in such a way that the bin content corresponding
to $\cos \theta^{*} = 0$ is equal to 1. The discrepancy in normalisation is
therefore hidden by this procedure.}, the theoretical prediction being
systematically above the data by a constant multiplicative factor $\sim 1.4$ 
for the scale choice used. 
One may wonder why, for the given scale choice in the NLO calculation, and for
the same data sample, the agreement in normalisation seems somewhat worse for
this distribution than what is found in average for the other infrared
insensitive distributions.

\vspace{0.2cm}

\noindent  
Firstly, it should be stressed that this $\cos \theta^{*}$ distribution is much
more tricky than its naming suggests. Indeed, the distribution shown in Fig.~\ref{fig4bis} is {\it not} literally the bare quantity $d \sigma/d \cos
\theta^{*}$. Let us recall that the `naive, bare' distribution $d \sigma/d \cos
\theta^{*}$, without additional cut, has a bell-like shape peaked at $y^{*} = 0$
and decreasing towards 0 as $|y^{*}|$ is increasing. This results from a
competition between the behaviour of the partonic matrix element squared, and
the fall-off of the pdf's when $x \to 1$. The matrix element squared behaves as as $(1 \pm \cos \theta^{*})^{-\alpha}$ when $\cos \theta^{*}$ increases, in the case of a massless spin $s$ exchange in the $t$ or $u$ channel: a
distribution unveiling this behaviour would namely probe the short distance
dynamics. On the other hand $x_{i} \to 1$ for $i = 1$ or $2$ when $|y^{*}|$
increases\footnote{See for instance \cite{owens-rmp} for explicit LO
expressions.}, and this effect wins. Additional cuts  are implemented to freeze
the influence of the pdf's and magnify the short distance dynamics. The
additional cuts restrict the data sample providing this $\cos \theta^{*}$
distribution: this sample is {\it not} the same as the one yielding the other
infrared insensitive distributions, but only a subsample of it. In particular,
an auxiliary selection cut on the invariant mass $m_{\pi \pi} \geq 7$ GeV is
imposed, in order to avoid $p_{T}$-threshold effects induced by the selection
cut $p_{T} > p_{T \, min}$ \cite{e706}. The whole distribution carved out this
way is dominated by the pairs with $m_{\pi \pi} \sim 7.0 - 8.0$ GeV. We then
notice on Fig.~\ref{fig2} for the invariant mass distribution that the ratio
data/NLO in the corresponding bins is approximately 0.7, a value consistent
with what is observed on Fig.~\ref{fig4bis} bottom for the $\cos \theta^{*}$
distribution. In summary the normalisation of the whole $\cos \theta^{*}$
distribution is given by the one of only a few bins of the $m_{\pi \pi}$ 
distribution. This makes the overall normalisation in the NLO $\cos \theta^{*}$ 
distribution very much scale dependent. If, instead of choosing the scale
`preferred' by the each pion's $p_{T}$ distribution in the theoretical
calculation as explained in sect. \ref{comparison-e706}, we take a scale 
which makes the ratio Data/Theory close to 1 for the $m_{\pi \pi}$ distribution, this apparent normalisation problem disappears.

\subsection{Infrared sensitive observables}\label{ir-sensitive}

We now successively examine the two families of infrared sensitive observables
which we identified at the beginning of this section. We first focus on the
distribution $d\sigma/d \phi_{\pi \pi}$ as an example of observables with strong
infrared sensitivity. Next we consider the cases of the $q_{T}$ and $Z$
distributions, as examples of observables for which the infrared sensitivity is 
smeared. 
Later we examine how the sensitivity of these types of observables to non
perturbative effects evolves with increasing individual transverse momenta of
the pions and/or with increasing invariant masses of the pairs. 

\subsubsection*{The angular distribution $d \sigma/d \phi_{\pi \pi}$}

As can be seen on Fig.~\ref{fig6}, the `NLO distribution' and the
experimental spectrum are absolutely different. The LO contribution to the `NLO
distribution' is proportional to $\delta(\pi - \phi_{\pi \pi})$, and the 
inelastic part of its NLO contribution diverges as ${\cal O}(\alpha_{s} \ln(\pi
- \phi_{\pi \pi})/(\pi - \phi_{\pi \pi}))$ when $\phi_{\pi \pi} \to \pi$.
Therefore the `NLO distribution' is mainly concentrated in the last bin
containing $\phi_{\pi \pi} = \pi$ at the elastic boundary of the spectrum. In
contrast the experimental distribution is spread out much more and flatter. The
discrepancy is so spectacular that one has to ask the question, why there is 
such a satisfactory agreement for the $p_T$ distribution, cf. Fig.~\ref{fig1}, 
and, similarly what the reason is for the sharp disagreement in the first two
bins for the invariant mass distribution in Fig.~\ref{fig2}, while the agreement
is satisfactory in the other bins. 

\vspace{0.2cm}

\noindent  
We notice that, although the shapes of the `NLO distribution' and of the
measured distribution are completely different, the area below these
distributions are approximately equal. This suggests that long distance
phenomena, which are ignored in the NLO calculation, proceed to nothing but a
drastic kinematic `reshuffle' of the configurations of the hard partons
involved in the subprocess into the ones of the measured pions, i.e. a
probabilistic redistribution of kinematic variables occurring with total
probability 1. This hypothesis is supported by the following argument. Looking
back to the invariant mass distribution of pairs, we noticed previously a
strong disagreement between the NLO prediction and the data in the first two
bins. Quantitatively, the theoretical result is more than a factor 25 smaller
than the measured central value for the second bin 4.5 GeV $\leq m_{\pi \pi}
\leq$ 5.0 GeV, and it is about a factor 60 smaller in the first bin 4.0 GeV
$\leq m_{\pi \pi} \leq$ 4.5 GeV. Given the fast fall-off of the $p_{T}$
distribution the pion pairs for which the individual pions' $p_{T}$ are
minimal i.e. close to 2.5 GeV both dominate the first bins of the invariant
mass distribution, and  control the $\phi_{\pi \pi}$ distribution. Furthermore, pairs with small $y^*$ dominate. Therefore, approximating
the identity (\ref{C}) by roughly $m_{\pi \pi} \sim 2 p_{T \, min}  \sin
(\phi_{\pi \pi}/2)$ we infer that the first and second bins in invariant mass
are dominated by pairs with respectively $\phi_{\pi \pi} \sim 115 \, ^{o}$ and
$\phi_{\pi \pi} \sim 140 \, ^{o}$. If we now look at the discrepancy between
the `NLO calculation' and the measured distribution for the $\phi_{\pi \pi}$
spectrum at these values of $\phi_{\pi \pi}$, we find respectively $\sim$ 40
and 20. These values are  comparable with the ones for the discrepancies
between NLO results and data for the invariant mass spectrum in the
corresponding first and second bin in invariant mass.

\vspace{0.2cm} 

\begin{figure}[htbp]
\begin{center}
\includegraphics[scale=0.7]{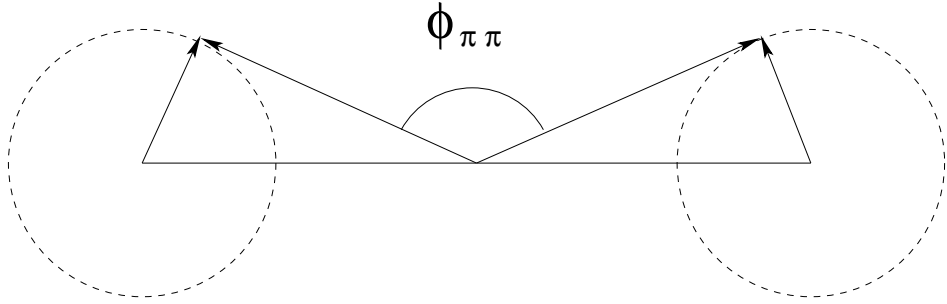}
\end{center}
\caption{\small Illustration of the reshuffle of back to back configurations by
final state $k_{T}$ kick. The radii of the circles represent the magnitude of the
kick. If the magnitude of the $k_{T}$ kick is not much smaller than the
selection cut $p_{T \, min}$, the $\phi_{\pi \pi}$ of the reshuffled 
configuration can be very different from $\pi$.
}
\label{thomas-circle3}
\end{figure}

\noindent 
A proper understanding of the shape of the distribution $d\sigma/d\phi_{\pi 
\pi}$ requires the all order summation of the large Sudakov logarithms which
appear order by order in the theoretical calculation. The latter come from soft
gluon emission in both initial and final states - and, beyond leading
logarithmic accuracy, from interferences between them. The colour structure
involved \cite{catani-colour,sterman-colour} is quite more intricate than  in
the well-known  processes of Drell-Yan \cite{ddk,parisi-petronzio,bcm,css} and
nearly back-to-back hadron production in hadronic $e^{+}e^{-}$ annihilation
\cite{cs}. This summation has not been performed yet  for the $\phi_{\pi \pi}$
distribution. Beyond this summation, non perturbative effects are also involved.
In the analysis of infrared sensitive distributions of electrodynamic probes in
fixed targets experiments, such as the transverse momentum distribution of
lepton pairs in Drell-Yan or of photon pairs in prompt photon pairs production
\cite{cfg-gamma}, such non perturbative effects have been shown to be important
over a large fraction of the spectrum. In the present case they may have even
larger effects. They may come in particular from some $k_{T}$ kick on initial
partons, as well as on the hadrons produced by the fragmentation process. Since
both the initial and final states are affected by these phenomena, and since
the selection cut on the $p_{T}$ of the pions is not much larger than the
typical scale $\sim {\cal O}(1$ GeV) of non perturbative phenomena in strong
interaction, even a rather moderate  effective $k_{T}$ kick - several hundreds
of MeV `per hard coloured leg' - may be able to reshuffle the hard parton
configuration drastically, as one can convince oneself with the cartoon drawn
in Fig.~\ref{thomas-circle3}. 

\vspace{0.2cm}

\noindent   
It is important to understand how it is conceivable that the infrared
insensitive observables can be reasonably estimated using perturbative, fixed
order calculations, whereas there exist distributions, such as 
$d\sigma/d\phi_{\pi \pi}$ which are so heavily distorted by non perturbative
effects. The key point is the following: if this probabilistic reshuffle acts
on kinematic variables which are integrated over during the calculation of some
distribution, and if the available phase space for this kinematic variable is
not severely amputated by some selection cut in the range considered, this
reshuffling leaves nearly no fingerprint in this observable. This seems to be
what happens for the distribution of each photon's $p_{T}$ (the variable
$\phi_{\pi \pi}$ is integrated out when the $\vec{p}_{T}$ of the other photon
is integrated out). The same conclusion holds also, for example, for the
invariant mass distribution as soon as the invariant mass is large enough
compared to $2 \, p_{T \, min}$. It does not hold when $m_{\pi \pi}$ is near or
below this value because of the restriction acting on $\phi_{\pi \pi}$ through
eqn.(\ref{C}).

\subsubsection*{Transverse momentum distribution of pairs, $d\sigma/dq_{T}$}

As we mentioned in the general introduction of this section, the situation in
the case at hand is different from the one usually faced concerning this 
observable in the Drell-Yan process/production of a weak vector boson, or in
the production of pairs of prompt photons\footnote{More precisely, we mean here
the production of photon pairs through the so-called `direct' mechanism, cf.
sect. \ref{theory} and Ref.~\cite{bgpw}. On the contrary the production of pairs
of `photons from fragmentation' is very similar to the present case.}. E.g., in
the Drell-Yan case, the calculation of this distribution is singular at $q_{T}
= 0$ order by order in perturbation theory. In particular the zeroth order
contribution to $d\sigma/dq_{T}^{2}$ is proportional to  $\delta(q_{T}^{2})$ by
transverse momentum conservation, and logarithmically enhanced `$+$' distributions are involved
at every successive order. On the contrary, in the present case, the double 
convolution with the two fragmentation functions makes the calculation finite at
every transverse momentum $q_{T}$ of the pair, order by order in perturbation
theory. In particular it is non vanishing already at lowest order for every $q_{T} > 0$ allowed by kinematics. One may therefore expect a somewhat reduced
infrared sensitivity, compared with the $\phi_{\pi \pi}$ distribution. The
comparison between the NLO calculation and the measured $q_{T}$ distribution,
cf. Fig.~\ref{fig8}, namely the examination of the
discrepancy between the former and the latter suggest that it is hardly the
case. Whereas the discrepancy reaches quantitatively a magnitude of 20 to 50 in
the $\phi_{\pi \pi}$ case, it is `only' about a factor 5 to 10 for the $q_{T}$
distribution. Moreover the NLO calculation is still inadequate in shape as well
over all the measured $q_{T}$. This inadequacy is related to the discrepancy in
the $\phi_{\pi \pi}$ distribution examined in the previous subsection. Let us
write $q_{T}^{2} = (p_{T1}-p_{T2})^{2} + 2 p_{T1}p_{T2} (1 + \cos \phi_{\pi
\pi})$. The $\phi_{\pi \pi}$ distribution is concentrated in the vicinity of
$\phi_{\pi \pi} = \pi$ whereas the experimental one is more flat and spread.
Therefore the NLO result for the $q_{T}$ is peaked near $q_{T} = 0$ and is
higher than the experimental spectrum in this region. As for configurations
with a larger $q_{T}$, they can have two origins. Either the individual
$p_{T}$'s have to be larger, but NLO and data agree on the $p_{T}$
distribution, hence this origin contributes by the same amount to the NLO
result and to the data; or $\phi_{\pi \pi}$ is far from $\pi$, in which case
the NLO $\phi_{\pi \pi}$ distribution is much below the experimental one. This
makes the NLO $q_{T}$ distribution fall faster than the measured one as $q_{T}$
gets larger. 

\subsubsection*{Transverse momentum balance distribution $d\sigma/dZ$}

This observable is another case of distribution which is singular e.g. in 
diphoton production, namely at the critical point $Z = 1$. Indeed, in diphoton
production, the contribution of the `direct' mechanism is
proportional to $\delta(1 - Z)$ at lowest order by transverse momentum
conservation, and involves logarithmic singularities at every higher order. Due
to the same reason as in the previous example, this distribution is, instead,
finite for all $Z$, order by order in perturbation theory. As can be seen on
Fig.~\ref{fig9}, the comparison between the NLO calculation
and the experimental distribution roughly agrees for $Z > 1$, but is
inadequate for $Z < 1$. This can be understood as follows.
From $Z = - p_{T1}/p_{T2} \cos \phi_{\pi \pi}$ one sees that $p_{T1} >
p_{T2}$  is a necessary condition to have $Z > 1$. Since this occurs preferably with values of $\phi_{\pi \pi}$ not far from $\pi$, one
understands that the NLO curve overshoots the data slightly regarding the
last bins of the $\phi_{\pi \pi}$ distribution, cf. Fig.~\ref{fig6}. Yet
the reversal of the ordering of the respective magnitudes of the NLO and
measured $\phi_{\pi \pi}$ distributions occurs for a value $\phi_{\pi \pi}$
close to $\pi$. The integration over the fragmentation variables averages
these two opposite orderings, and makes the discrepancy between the NLO and
measured $Z$ distribution rather small for $Z > 1$. On the other hand, for
$Z < 1$, $p_{T1}$ and $p_{T2}$ are unconstrained and most likely of
comparable size. This $Z$ region thus selects the lower range of $\cos
\phi_{\pi \pi}$  and for small $Z$  one finds a similar large discrepancy
as for the $\phi_{\pi \pi}$ distribution in the lower $\phi_{\pi \pi}$ range.

\subsection{Energy dependence of the non perturbative effects}\label{evol}

The sensitivity of the infrared sensitive observables examined above to non
perturbative aspects is strong and extended to a large fraction, if not the
totality, of the measured spectrum. This is mainly because the E706 data  were
selected above a cut $p_{T\, min} \geq 2.5$ GeV, and because the normalisations
and shapes of these distributions are dominated by the pairs in which the
individual $p_{T}$'s of the pions are close to $p_{T \, min}$, a value not
large with respect to the typical scale of non perturbative effects. As we are
interested in making predictions for collider experiments, especially for the
LHC \cite{bgpw2}, it would be interesting to see how the situation evolves with the scales. Two situations have to be distinguished: 1) the increase of 
the `hard scale', loosely speaking -- i.e. the invariant mass of the pair, or
the individual $p_{T}$'s of the pions in pairs -- at fixed
$\sqrt{S}$, and 2) the increase of $\sqrt{S}$ at fixed `hard scale'. 

\vspace{0.2cm}

\noindent    
In the archetype of the Drell-Yan process, for the distribution 
$d\sigma/dq_{T} dm$ of lepton pairs, this question has been studied relying on
resummed calculations. In particular, the authors of
Refs.~\cite{parisi-petronzio,css} have studied how the sensitivity to non
perturbative effects at $q_{T} = 0$ evolves as a function of $m$. They
determined which region dominates the Sudakov-type factor in impact parameter
($b$) space, by mean of a saddle point analysis. They showed that the dominant
region in $b$ space is purely non perturbative at $m$ values in the fixed
target energy range, but that it migrates towards smaller values of $b$ and
ultimately fits completely inside the perturbative region $\Lambda_{QCD} b \ll
1$ when $m$ becomes large enough. This indicates that the non perturbative
effects become confined to a closer and closer vicinity of the critical point
as the invariant mass $m$ increases. This feature makes it difficult to extract
a precise parametrisation of these non perturbative effects from experimental
data \cite{dws,ladinsky-yuan,erv}. On the other hand it implies that 
theoretical predictions at higher, collider energies are less affected.

\vspace{0.2cm}

\noindent
In this subsection, we perform a phenomenological study through which we try to
extract information about item 1) within the range covered by the E706
experiment. As noticed in the last two subsections, the $\phi_{\pi \pi}$
distribution is the key of the understanding of the others. A useful
information could be extracted from data samples ordered according to
increasing values of $p_{T \, min}$ for fixed $\sqrt{S}$. Tabulations of data
of this type are available for the invariant mass distribution. However not
much information can be learned from it since this observable is blind to non
perturbative effects over most of its spectrum. On the other hand, such
tabulations are not available for the more instructive $\phi_{\pi \pi}$
distribution. An alternative could be provided by a refined double binning of
the data according to the infrared sensitive variable considered and,
simultaneously, to increasing invariant mass, the latter corresponding loosely
to increasing individual $p_{T}$'s of each pion in pairs. This type of double
binning is available for the $p_{T \, out}$ distribution, but again not for the
$\phi_{\pi \pi}$ distribution. The only information available about $\phi_{\pi
\pi}$ in this respect is the average $<\!\phi_{\pi \pi}\!>$ \cite{e706}. 

\subsubsection*{Increasing the hard scale at fixed $\sqrt{S}$.}

\noindent
Let us thus consider the average values of the azimuthal angle difference
between the two pions of pairs, $<\!\phi_{\pi \pi}\!>$ as a function of the
invariant mass of the pair. Of course the shape of the $\phi_{\pi \pi}$
distribution is infrared sensitive near $\phi_{\pi \pi} = \pi$ and, to any
fixed order in perturbation theory, the theoretical estimate of the $\phi_{\pi
\pi}$ distribution is divergent at $\phi_{\pi \pi}= \pi$. However when the
average $<\!\phi_{\pi \pi}\!>$ is computed, the Sudakov logarithms are smeared out
by the integration order by order in perturbation theory. There is no
logarithmic enhancement in $<\!\phi_{\pi \pi}\!>$ order by order in perturbation
theory anymore. Consequently, a fixed order calculation of $<\!\phi_{\pi \pi}\!>$
is expected to become more and more accurate for harder and harder pion pairs,
without a detailed knowledge of the shape. Non perturbative effects are
expected to induce deviations in the form of power corrections 
${\cal O} \left( 1 \, \mbox{GeV}/p_{T} \right)$, 
${\cal O} \left( 1 \, \mbox{GeV}/m_{\pi\pi} \right)$.

\vspace{0.2cm}

\begin{figure}[htbp]
\begin{center}
\includegraphics[width=0.8\linewidth,height=12 cm]{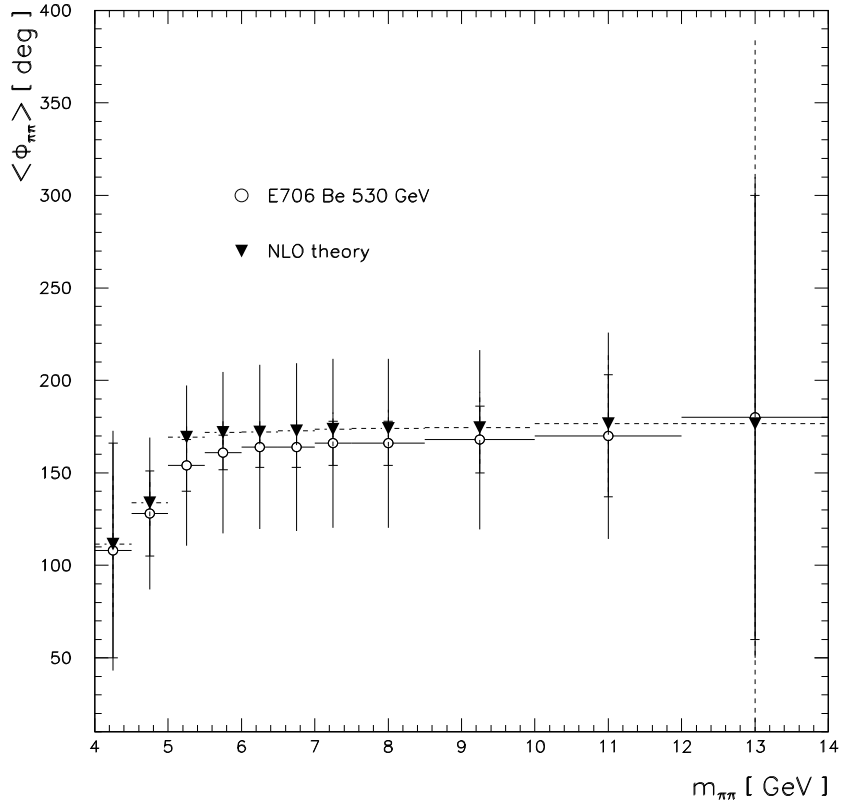}
\end{center}
\caption{\small Dipion distribution $<d\phi_{\pi \pi}>/dm_{\pi \pi}$ vs. 
 the invariant mass $m_{\pi \pi}$, in $p-Be$ 
 collisions with a beam energy of $530$ GeV.
 Data points with statistical and systematic errors in quadrature 
 are from the E706 collaboration ~\protect\cite{e706}.
 The NLO prediction with scales $M = \mu = M_{f} = 3/8 \, (p_{T1} + p_{T2})$
 is shown as triangles.
 } 
\label{fig-fimean}
\end{figure}

\noindent
For $m_{\pi \pi} > 5$ GeV, the full range $105 ^{o} \leq \phi \leq 180 ^{o}$ is
accessible\footnote{The situation in the first bins with $m_{\pi \pi} \leq 5$
GeV is peculiar. In these bins the selection cut $p_{T}> p_{T \, min}$ forbids
the back-to-back region as discussed previously. The average $<\!\phi_{\pi \pi}\!>$
is thus much smaller than $\pi$. This holds both for the NLO result and the 
data, and no relevant information can be extracted regarding the problem
concerned by this discussion.},  and $<\!\phi_{\pi \pi}\!>$ is dominated by the
nearly back-to-back region. This dominance of the region $\phi_{\pi \pi} \sim
\pi$ is actually  stronger in the NLO result than in the data. Indeed, the
pairs of pions with low individual $p_{T}$ and relatively low invariant mass
dominate the $\phi_{\pi \pi}$ distribution discussed in subsect.
\ref{ir-sensitive}. The latter is thus strongly sensitive to non perturbative
effects, which make it much flatter than the shape of the NLO result. Therefore
the experimental values of $<\!\phi_{\pi \pi}\!>$ should be lower than the NLO
values. The distribution $d\sigma/dm_{\pi \pi}d\phi_{\pi \pi}$ should then
become steeper and more concentrated near the back-to-back region with
increasing invariant mass, if non perturbative effects become less and less
important. The experimental values of $<\!\phi_{\pi \pi}\!>$ should increase with
increasing invariant mass, and the difference between them and the
corresponding NLO results should decrease. The general trend of the data seem
to agree with this expectation, as suggested by Fig.~\ref{fig-fimean}. To be 
fair
however, the very large error bars prevent from drawing any firm conclusion.
More generally, the first moment of infrared sensitive distribution is not
sensitive enough to long distance effects. Cumulants of higher orders of these
distributions would be much more instructive in this respect, not speaking about
these distributions themselves.

\vspace{0.2cm}

\begin{figure}[htbp]
\begin{center}
\includegraphics[width=0.8\linewidth,height=12 cm]{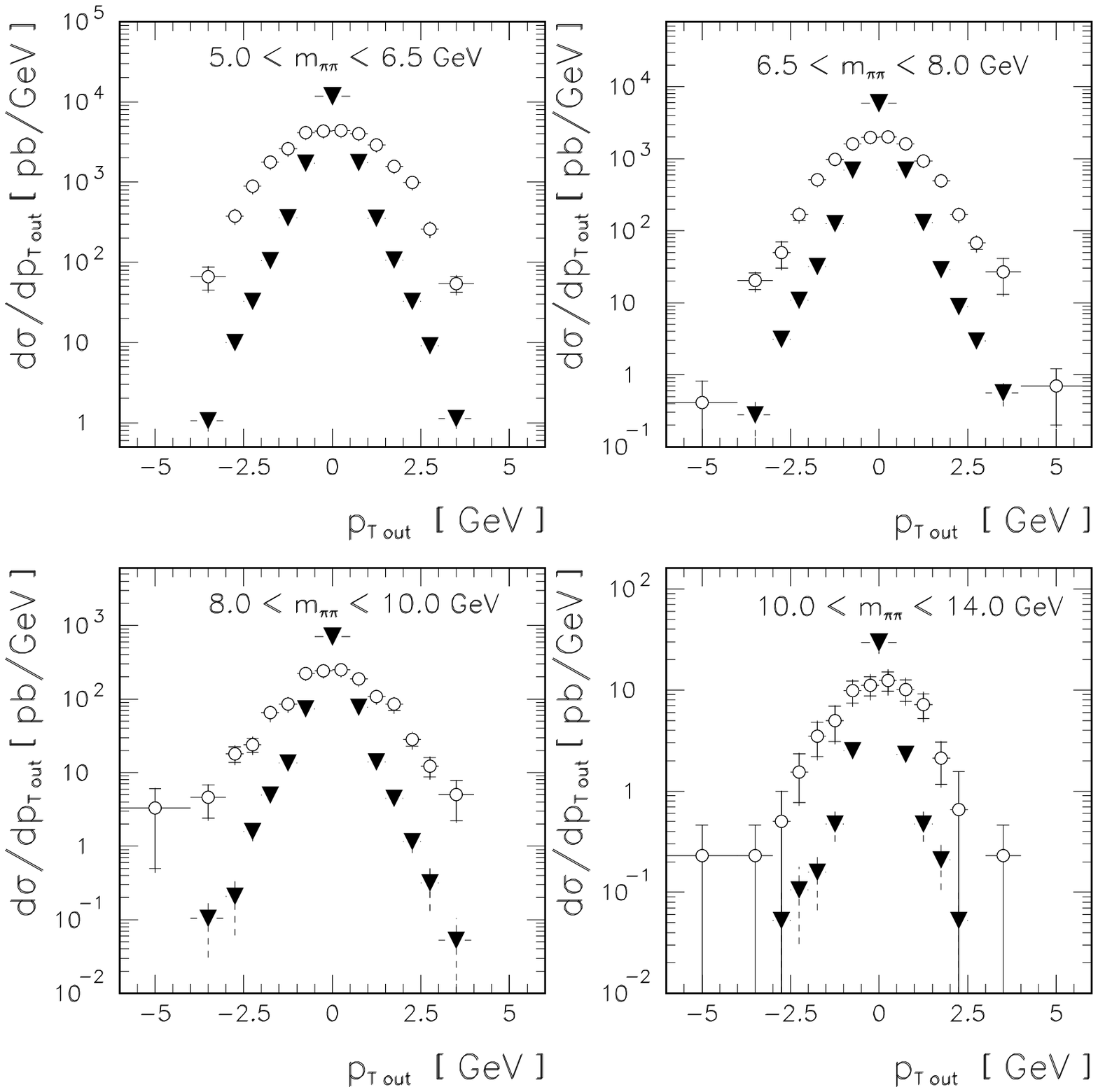}
\end{center}
\caption{\small Dipion differential cross section Dipion differential cross 
 section $d\sigma/dp_{T \, out}$ vs. $p_{T \, out}$ in $p-Be$ collisions with 
 a beam energy of $530$ GeV. Here $d\sigma/dp_{T \, out}$ stands for
 $\int dm_{\pi \pi} (d\sigma/dp_{T \, out} dm_{\pi \pi})$
 integrated over the following invariant mass slices: 
  $5.0 < m_{\pi \pi} <  6.5$ GeV (top left), 
  $6.5 < m_{\pi \pi} <  6.5$ GeV (top right),
  $8.0 < m_{\pi \pi} < 10.0$ GeV (bottom left) and 
 $10.0 < m_{\pi \pi} < 14.0$ GeV (bottom right).
 Data points with statistical and systematic errors in quadrature 
 are from the E706 collaboration ~\protect\cite{e706}.
 The NLO prediction with scales $M = \mu = M_{f} = 3/8 \, (p_{T1} + p_{T2})$
 is shown as triangles.
 } 
\label{fig-pout-mgg}
\end{figure}

\begin{figure}[htbp]
\begin{center}
\includegraphics[width=0.8\linewidth,height=12 cm]{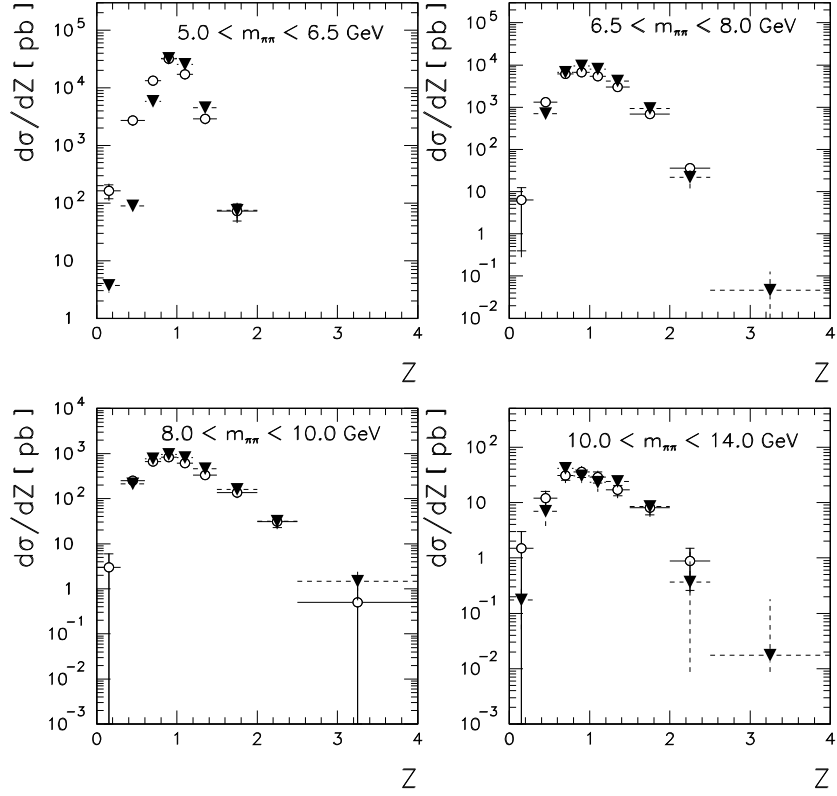}
\end{center}
\caption{\small Dipion differential cross 
 section $d\sigma/dZ$ vs. $Z$ in $p-Be$ collisions with a beam energy of 
 $530$ GeV. Here $d\sigma/dZ$ stands for
 $\int dm_{\pi \pi} (d\sigma/dZ dm_{\pi \pi})$
 integrated over the following invariant mass slices: 
  $5.0 < m_{\pi \pi} <  6.5$ GeV (top left), 
  $6.5 < m_{\pi \pi} <  6.5$ GeV (top right),
  $8.0 < m_{\pi \pi} < 10.0$ GeV (bottom left) and 
 $10.0 < m_{\pi \pi} < 14.0$ GeV (bottom right).
 Data points with statistical and systematic errors in quadrature 
 are from the E706 collaboration ~\protect\cite{e706}.
 The NLO prediction with scales $M = \mu = M_{f} = 3/8 \, (p_{T1} + p_{T2})$
 is shown as triangles.
 } 
\label{fig-z-mgg}
\end{figure}

\noindent
We examined also the distribution $d\sigma/dp_{T \, out}dm_{\pi \pi}$ sliced in
broad bins of invariant mass. As seen on Fig.~\ref{fig-pout-mgg} all plots for
each separate bin of invariant mass are similar to the one shown in Fig.~\ref{fig7}. The NLO result is inadequate to account for the measured
distribution for all invariant masses, and nothing instructive can be extracted
from it regarding the range of influence of non perturbative effects with
respect to the invariant mass. Last, in the case of the distribution
$d\sigma/dZdm_{\pi \pi}$, we notice on Fig.~\ref{fig-z-mgg} that the
discrepancy between the NLO result and the data for $Z < 1$ decreases with
increasing $m_{\pi \pi}$. This holds at least for $Z$ values not very small
with respect to 1, for bins in invariant mass which gather enough statistics so
that any conclusion can be drawn.

\vspace{0.2cm}

\noindent
It would have been interesting also to identify how the magnitude of non
perturbative effects evolves when the beam energy increases from 530 to 800 GeV. 
Unfortunately, the level arm in $\sqrt{S}$ is too small and the errors bars 
are too large to extract this information.

\section{Conclusions and outlook}\label{concl}

We have compared NLO predictions with data of the E706 experiment for various
observables. We have classified the observables according to their respective
infrared sensitivity. For the class of infrared insensitive observables the NLO
predictions agree rather well with the corresponding measured ones over the whole
measured spectrum.  In contrast, the NLO result is inadequate to describe
infrared sensitive observables, not only in the close vicinity of the infrared
sensitive region as expected, but even over a large fraction of, if not the whole
spectrum. The phenomenological analysis performed in this article suggests that
the discrepancy can be understood in terms of a probabilistic reshuffle of the
partonic configurations involved in the NLO calculation.  From a theoretical
point of view a Sudakov-type summation of logarithmically enhanced infrared terms
which appear order by order in perturbation theory in the expansion of these
infrared sensitive observables is mandatory. Yet it is likely to be not enough, as the discrepancy is probably dominated by large
non perturbative effects.  The non perturbative dynamics strongly distorts the
infrared sensitive observables because the experimental selection cut $p_{T \,
min}$ imposed on each pion is not large compared to the typical non perturbative
scale ${\cal O}(1)$ GeV. On the other hand, these non perturbative
effects turn out to be weak on infrared insensitive distributions. In particular
they are much weaker than in the case of one particle inclusive production,
especially on the $p_{T}$ distribution in the lower $p_{T}$ range. This can be
traced back to the absence of trigger bias in pair production when symmetric cuts
are applied, contrarily to the situation  in one particle production. A further
examination of the E706 data tends to suggest that these non perturbative effects
become less and less important in the infrared sensitive distributions, and in
particular more and more confined to the vicinity of the critical points, as the
$p_{T}$'s of the selected pions increase for fixed $\sqrt{S}$, although, to be
fair, no definite conclusion can be extracted from this examination due to large
experimental uncertainties. On the other hand no evolution of the magnitude of
non perturbative effects can be seen when going from 530 to 800 GeV because the
level arm in $\sqrt{S}$ is not large enough.

\vspace{0.2cm}

\noindent
We expect these features to hold as we move from fixed target
to collider experiments, as soon as the selection cut $p_{T \, min}$ increases
up to 10-15 GeV at the Fermilab Tevatron and 20-40 GeV at the forthcoming LHC.
We therefore expect NLO predictions to be reliable in particular for inclusive
observables concerning pions pairs - as well as $\gamma \pi^{0}$ pairs - at the
Tevatron and LHC, at least far enough from  the boundaries of the spectrum. On
the other hand, near these boundaries, the  various kinematic cuts can induce
extra infrared sensitivity in partonic calculations. This is certainly
expected to hold  for observables which do not require any isolation cut.
On the other hand, the implementation of isolation can affect the ability of
fixed order perturbative calculations to provide reliable predictions for most
inclusive distributions. This problem, which is of particular relevance in the
calculation of backgrounds to the Higgs boson search at LHC in the channel $H
\to \gamma \gamma$, will be examined in detail in a forthcoming article
\cite{newcitation}.

\vspace{0.2cm}

\noindent 
{\bf Acknowledgments.} We wish to thank M. Zelinski for several useful
discussions and M. Begel for correspondence. We thank also M. Fontannaz 
for relevant remarks in the comparison between one particle inclusive and pair
production. This work was supported in part by
the EU Fourth Training Programme ``Training and Mobility of Researchers",
Network ``Quantum Chromodynamics and the Deep Structure of Elementary
Particles", contract FMRX-CT98-0194 (DG 12 - MIHT). LAPTH is a ``Unit\'e Mixte
de Recherche  (UMR 5108) du CNRS associ\'ee \`a l'Universit\'e de Savoie".

\vspace{0.2cm} 

\noindent
{\bf Note added.} During the completion of this article, we became aware of a
recently submitted article by J.F. Owens \cite{owens-dihad} also dedicated to 
dihadron production.

\newpage

\section*{Figures}
 
\pagebreak

\end{document}